\documentclass[]{jfm}

\usepackage{graphicx}
\usepackage{newtxtext}
\usepackage{newtxmath}
\usepackage{natbib}
\usepackage{hyperref}
\usepackage{comment}
\usepackage[usenames, dvipsnames]{xcolor}
\usepackage{amssymb, latexsym}
\usepackage{amsmath}
\usepackage{mathrsfs}
\usepackage{soul}
\usepackage{amsbsy}
\usepackage{psfrag}
\usepackage{mathtools}
\usepackage{color}
\usepackage{subfig}
\usepackage{makecell}
\usepackage[makeroom]{cancel}
\usepackage{tikz}
\usepackage{tikz-3dplot}
\usetikzlibrary{decorations.markings,arrows.meta,calc,shapes,intersections,backgrounds,patterns.meta,bending,angles,perspective,decorations.pathreplacing,decorations.pathmorphing}
\usepackage{xifthen}
\usepackage{gacaps}
\usepackage[normalem]{ulem}
\usepackage{pgfplots}
\usepgfplotslibrary{fillbetween}
\pgfplotsset{compat=1.12} 

\definecolor{myBlue}{rgb}{0, 0.4470, 0.7410}
\definecolor{myRed}{rgb}{0.8500, 0.3250, 0.0980}
\hypersetup{
    colorlinks = true,
    urlcolor   = blue,
    citecolor  = black,
}

\newcommand{\RomanNumeralCaps}[1]
\linenumbers

\let\tw\textwidth

\def\dd{{\, \rm{d}}}

\def\beq{\begin{equation}}
\def\eeq{\end{equation}}

\def\dd{{\, \rm{d}}}

\def\beq{\begin{equation}}
\def\eeq{\end{equation}}

\def\dis{\varepsilon}

\definecolor{olivegreen}{rgb}{0,0.6,0}

\def\drawline#1#2{\raise 2.5pt\vbox{\hrule width #1pt height #2pt}}

\def\trian{\raise 1.25pt\hbox{$\scriptstyle\triangle$}\nobreak}

\def\dtrian{\raise 1.25pt\hbox%
{$\scriptscriptstyle\bigtriangledown$}\nobreak}

\def\squar{\raise 1.25pt\hbox{$\scriptstyle\Box$}\nobreak}

\def\diamon{\raise 1.25pt\hbox{$\scriptstyle\diamond$}\nobreak}

\def\dd{{\, \rm{d}}}

\def\beq{\begin{equation}}
\def\eeq{\end{equation}}

\definecolor{C0}{HTML}{1F77B4} 
\definecolor{C1}{HTML}{FF7F0E}
\definecolor{C2}{HTML}{2CA02C}
\definecolor{C3}{HTML}{D62728}
\definecolor{C4}{HTML}{9467BD}
\definecolor{C5}{HTML}{8C564B}
\definecolor{C6}{HTML}{E377C2}
\definecolor{C7}{HTML}{7F7F7F}
\definecolor{C8}{HTML}{BCBD22}
\definecolor{C9}{HTML}{17BECF}

\let\bs\boldsymbol
\let\provc\providecommand
\newcommand{\mybq}{\bs{q}}        
\newcommand{\bq}{\bs{q}}        
\newcommand{\bQ}{\bs{Q}}        
\newcommand{\bi}{\bs{i}}        
\newcommand{\bj}{\bs{j}}        
\newcommand{\Is}{\tilde{\imath}}        
\newcommand{\Iss}{\zeta}                
\newcommand{\GIs}{\tilde{\mathcal{G}}}  

\newcommand{\myindexvar}[3]{%
  \ifthenelse{\isempty{#2}\and\isempty{#3}}%
  {#1}{#1_{#2}^{#3}}} 

\provc{\sq}[2]{\myindexvar{q}{#1}{#2}}       
\provc{\bsq}[2]{\myindexvar{\mybq}{#1}{#2}}  

\newcommand{\mybS}{\bs{S}}        
\newcommand{\mybA}{\bs{A}}        
\newcommand{\mybth}{\bs{\theta}}  
\newcommand{\mybJ}{\bs{J}}        

\provc{\bS}[2]{\myindexvar{\mybS}{#1}{#2}}     
\provc{\bA}[2]{\myindexvar{\mybA}{#1}{#2}}     
\provc{\bth}[2]{\myindexvar{\mybth}{#1}{#2}}   

\provc{\bWs}[2]{\myindexvar{\bs{W}}{#1}{#2}}               
\provc{\bWa}[2]{\myindexvar{\bs{V}}{#1}{#2}}   

\provc{\bJ}[2]{\myindexvar{\mybJ}{#1}{#2}}                  
\provc{\bJtilde}[2]{\myindexvar{\widetilde{\mybJ}}{#1}{#2}} 
\provc{\bJhat}[2]{\myindexvar{\widehat{\mybJ}}{#1}{#2}}     

\provc{\J}[2]{\myindexvar{J}{#1}{#2}}                  
\provc{\Jtilde}[2]{\myindexvar{\widetilde{J}}{#1}{#2}} 
\provc{\Jhat}[2]{\myindexvar{\widehat{J}}{#1}{#2}}     

\provc{\mun}{\mu}
\provc{\sgn}{\Xi}
\provc{\bmun}{\bs{\mu}}
\provc{\bsgn}{\bs{\Xi}}
\provc{\muopt}{\mu^*}
\provc{\sgopt}{\Xi^*}
\provc{\bmuopt}{\bs{\mu}^*}
\provc{\bsgopt}{\bs{\Xi}^*}
\provc{\mutar}{\hat{\mu}}
\provc{\sgtar}{\widehat{\Xi}}
\provc{\bmutar}{\hat{\bs{\mu}}}
\provc{\bsgtar}{\widehat{\bs{\Xi}}}

\provc{\relf}{\alpha}
\provc{\relfmu}{\relf_\mu}
\provc{\relfsg}{\relf_\xi}

\provc{\bths}{\mybth_s} 
\provc{\bthpa}{\mybth_p}
\provc{\bthaa}{\mybth_a}

\provc{\ths}{\theta_s} 
\provc{\thpa}{\theta_p}
\provc{\thaa}{\theta_a}

\provc{\shiftM}{\bs{a}}
\provc{\shiftmi}{a}

\definecolor{myc1}{HTML}{003049}
\definecolor{myc2}{HTML}{d62828}
\definecolor{myc3}{HTML}{f77f00}
\definecolor{myc4}{HTML}{6ca13b}

\newcommand{\myls}[2]{\lineSymbol[#1]{none}{#2}{.8ex}{.8pt}{#2}{#2}}

\title{Information-theoretic causality and applications to turbulence: energy cascade and inner/outer layer interactions}
\author{Adri\'an Lozano-Dur\'an\corresp{\email{adrianld@mit.edu}}, Gonzalo Arranz \and Yuenong Ling}
\affiliation{Department of Aeronautics and Astronautics, Massachusetts Institute of Technology, Cambridge, MA 02139, USA}

\begin{document}

\maketitle 



\begin{abstract}
We introduce an information-theoretic method for quantifying causality
in chaotic systems. The approach, referred to as IT-causality,
quantifies causality by measuring the information gained about future
events conditioned on the knowledge of past events.  The causal
interactions are classified into redundant, unique, and synergistic
contributions depending on their nature.  The formulation is
non-intrusive, invariance under invertible transformations of the
variables, and provides the missing causality due to unobserved
variables. The method only requires pairs of past-future events of the
quantities of interest, making it convenient for both computational
simulations and experimental investigations.  IT-causality is
validated in four scenarios representing basic causal interactions
among variables: mediator, confounder, redundant collider, and
synergistic collider. The approach is leveraged to address two
questions relevant to turbulence research: i) the scale locality of
the energy cascade in isotropic turbulence, and ii) the interactions
between inner and outer layer flow motions in wall-bounded
turbulence. In the former case, we demonstrate that causality in the
energy cascade flows sequentially from larger to smaller scales
without requiring intermediate scales. Conversely, the flow of
information from small to large scales is shown to be redundant. In
the second problem, we observe a unidirectional causality flow, with
causality predominantly originating from the outer layer and
propagating towards the inner layer, but not vice versa. The
decomposition of IT-causality into intensities also reveals that the
causality is primarily associated with high-velocity streaks. 
The Python scripts to compute IT-causality can be found
\href{https://github.mit.edu/Computational-Turbulence-Group/itcausalitytools}{here}.
\end{abstract}



\section{Introduction}
\label{sec:introduction}

Causality is the mechanism through which one event contributes to the
genesis of another~\citep{pearl2009}. Causal inference stands as a
cornerstone in the pursuit of scientific knowledge~\citep{bunge2017}:
it is via the exploration of cause-and-effect relationships that we
are able to gain understanding of a given phenomenon and to shape the
course of events by deliberate actions.  Despite its ubiquity, the
adoption of specialized tools tailored for unraveling causal
relationships remains limited within the turbulence community.  In
this work, we propose a non-intrusive method for quantification of
causality formulated within the framework of information theory.

Causality in turbulence research is commonly inferred from a
combination of numerical and experimental data interrogation. The
tools encompass diverse techniques such as statistical analysis,
energy budgets, linear and nonlinear stability theory, coherent
structure analysis, and modal decomposition, to name a
few~\citep[e.g.][]{robinson1991, reed1996, cambon1999, schmid2007,
  smits2011, jimenez2012, kawahara2012, mezic2013, haller2015,
  wallace2016, jimenez2018, marusic2019}.  Additional research
strategies involve the use of time cross-correlation between pairs of
time signals representing events of interest as a surrogate for causal
inference. For instance, investigations into turbulent kinetic
energy~\citep{jimenez2018, cardesa2015} and the spatiotemporal aspects
of spectral quantities~\citep[e.g.,][]{choi1990, wallace2014,
  wilczek2015, kat2015, he2017, wang2020} exemplify some of these
efforts. However, it is known that correlation, while informative,
does not inherently imply causation as it lacks the directionality and
asymmetry required to quantify causal interactions~\citep{beebee2012}.
While the aforementioned tools have significantly advanced our
physical understanding of turbulence, extracting rigorous causal
relationships from the current methodologies remains a challenging
task.

An alternative and intuitive definition of causality is rooted in the
concept of interventions: the manipulation of the causing variable
results in changes in the effect~\citep{pearl2009,
  eichler2013}. Interventions provide a pathway for evaluating the
causal impact that one process $A$ exerts on another process $B$. This
is achieved by adjusting $A$ to a modified value $\widetilde{A}$ and
subsequently observing the post-intervention consequences on
$B$. Within the turbulence literature, there are numerous examples
where the equations of motion are modified to infer causal
interactions within the system~\citep[e.g.][to name a
  few]{jimenez1991, jimenez1999, hwang2010, farrell2017b,
  lozano2021a}. Despite the intuitiveness of interventions as a
measure of causality, this approach is not without
limitations~\citep{eberhardt2007}. Causality with interventions is
intrusive (i.e., it requires modifying the system) and costly
(simulations need to be recomputed for numerical experiments). When
data is gathered from physical experiments, establishing causality
through interventions can be even more challenging or impractical (for
example, in a wind tunnel setup). Additionally, the concept of
causality with interventions prompts questions about the type of
intervention that must be introduced and whether this intervention
could impact the outcome of the exercise as a consequence of forcing
the system out of its natural attractor.

The framework of information theory, the science of message
communication~\citep{shannon1948}, provides an alternative,
non-intrusive definition of causality as the information transferred
from the variable $A$ to the variable $B$. The origin of this idea can
be traced back to the work of \citet{wiener1956}, and its initial
quantification was presented by \citet{granger1969} through signal
forecasting using linear autoregressive models. In the domain of
information theory, this definition was formally established by
\citet{massey1990} and \citet{kramer1998} through the utilization of
conditional entropies, employing what is known as directed
information. Building upon the direction of information flow in Markov
chains, \citet{schreiber2000} introduced a heuristic definition of
causality termed transfer entropy. In a similar vein,
\citet{liang2006} and subsequently \citet{sihna2016} suggested
inferring causality by measuring the information flow within a
dynamical system when one variable is momentarily held constant for
infinitesimally small times.  More recently, \citet{Lozano2022}
proposed a new information-theoretic quantification of causality for
multivariate systems that generalizes the definition of transfer
entropy. Other methods have been proposed for causal discovery beyond
the framework of information theory. These include nonlinear
state-space reconstruction based on Takens'
theorem~\citep{arnhold1999, sugihara2012}, conditional
independence-based methods~\citep{runge2018a, runge2018b}, restricted
Perron-Frobenius operator~\citep{jimenez2023}, reconstruction of
causal graphs and Bayesian networks~\citep{rubin1974, spirtes2000,
  pearl2000, koller2009, imbens2015}. \citet{runge2019} offers an
overview of the methods for causal inference in the context of Earth
system sciences. The reader is also referred to \citet{camps2023} for
a comprehensive summary of causality tools across disciplines.

The development and application of causality tools in the realm of
turbulence research remain scarce. Among the current studies, we can
cite the work by \citet{Tissot2014}, who used Granger causality to
investigate the dynamics of self-sustaining wall-bounded
turbulence. \citet{materassi2014} used normalized transfer entropy to
study the cascading process in synthetic turbulence generated via the
shell model.  \citet{liang2016} and \citet{lozano2019b} applied
information-theoretic definitions of causality to unveil the dynamics
of energy-containing eddies in wall-bounded turbulence.  A similar
approach was followed by \citet{wang2021} and \citet{wang2022} to
study cause-and-effect interactions in turbulent flows over porous
media.  \citet{Lozano2022} used information flux among variables to
study the energy cascade in isotropic turbulence.  Recently,
\citet{Martinez2023} used transfer entropy to analyze the formation
mechanisms of large-scale coherent structures in the flow around a
wall-mounted square cylinder. The new method proposed here to quantify
causality addresses important deficiencies compared to existing tools
in the literature.

This work is organized as follows. First, we discuss key aspects of
causal inference in \S\ref{sec:basic_causal}. The method is introduced
in \S\ref{sec:method}. The fundamentals of information theory required
to formulate the problem of IT-causality are presented in
\S\ref{sec:fundamentals} and the method is introduced in
\S\ref{subsec:infocau} along with a simple example and a discussion of
its properties. The approach is validated in simple stochastic systems
in \S\ref{sec:validation}.  IT-causality is used to investigate two
important problems in turbulence: the locality in scale of the energy
cascade (\S\ref{sec:application_cascade}); and the interaction between
flow motions in the inner layer and outer layer of wall-bounded
turbulence (\S\ref{sec:application_inner-outer}). Finally, limitations
of the method are outlined in \S\ref{sec:limitations} and conclusion
are offered in \S\ref{sec:conclusions}.

\subsection{Key concepts in causal inference}
\label{sec:basic_causal}

We discuss the distinction between causality, association, and
correlation, as well as descriptions of the three basic categories of
interactions between variables: mediator, confounder, and
collider. These are well-established concepts in the field of causal
inference~\citep[see, for example][]{pearl2009}, yet they remain
relatively underutilized within the turbulence research community.

Causality refers to the process by which one variable directly
influences or determines changes in another variable. In causal
relationships, a modification in the cause leads to a corresponding
alteration in the effect, and there is a logical and temporal
connection between them. For example, the rainy season in a particular
region is a cause of increased umbrella sales. On the other hand,
association signifies a statistical connection between two variables
where they tend to occur together more frequently than expected by
chance. Association does not necessarily imply causation; it could
result from common causes, statistical coincidences, or confounding
factors. One example of association is the increase in umbrella sales
and raincoat sales, which tends to happen concurrently due to the
confounding factor of the rainy season. Correlation is a specific form
of association that quantifies the strength and direction of a
linear~\citep{pearson1895} or monotonic~\citep{Spearman1987}
relationship between two variables. Correlation does not imply
causation and can even fail to identify associations. In general,
correlation implies association but not causation. Causation implies
association but not correlation~\citep{Altman2015}.

\begin{figure}
  \begin{center}
    \begin{tikzpicture}[cir/.style={circle,draw=black!70!white,,thick,inner sep=.5em},
    >={Latex[length=.2cm]}]
    
    \colorlet{cA}{myc2!30!white}
    \colorlet{cB}{myc1!30!white}
    \colorlet{cC}{myc3!30!white}

    \node [cir,fill=cB] (B) at (0,0)   {$B$}; 
    \node [cir,fill=cA] (A) at (-1,-1) {$A$}; 
    \node [cir,fill=cC] (C) at (+1,-1) {$C$}; 

    \path[thick,->]
    (A) edge[bend left] node [right] {} (B)
    (B) edge[bend right] node [left] {} (C);

    \node[anchor=north] at (0,-1.5) {(a) Mediator};

    \begin{scope}[xshift=.3\tw]
    \node [cir,fill=cB] (B) at (0,0)   {$B$}; 
    \node [cir,fill=cA] (A) at (-1,-1) {$A$}; 
    \node [cir,fill=cC] (C) at (+1,-1) {$C$}; 

    \path[thick,->]
    (B) edge[bend left] node [right] {} (A)
    (B) edge[bend right] node [right] {} (C);

    \node[anchor=north] at (0,-1.5) {(b) Confounder};
    \end{scope}

    \begin{scope}[xshift=.6\tw]
    \node [cir,fill=cB] (B) at (0,0)   {$B$}; 
    \node [cir,fill=cA] (A) at (-1,-1) {$A$}; 
    \node [cir,fill=cC] (C) at (+1,-1) {$C$}; 

    \path[thick,->]
    (A) edge[bend left] node [right] {} (B)
    (C) edge[bend right] node [right] {} (B);

    \node[anchor=north] at (0,-1.5) {(c) Collider};
    \end{scope}

\end{tikzpicture}
 \end{center}
\caption{ Schematic representation of mediator, confounder, and
  collider variables. \label{fig:basics_causality}}
\end{figure}
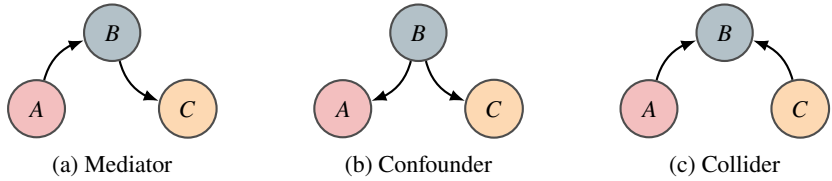

Another crucial factor to consider is the manner variables
interact. Let us consider three variables: $A$, $B$, and $C$. We can
identify three fundamental types of interactions as summarized in
figure~\ref{fig:basics_causality}.
\begin{itemize}
\item \emph{Mediator variables} emerge in the causal chain from $A$ to
  $C$, with the variable $B$ acting as a bridge: $A\rightarrow B
  \rightarrow C$. In this scenario, $B$ is often viewed as the
  mechanism or mediator responsible for transmitting the influence of
  $A$ to $C$. Mediator variables help explain the underlying
  mechanisms by which an independent variable influences a dependent
  variable. A simple example is $\uparrow$ education level
  $\rightarrow$ $\uparrow$ job skills $\rightarrow$ $\uparrow$ income.
\item \emph{Confounder variables} serve as a common cause for two
  variables: $B \rightarrow A$ and $B \rightarrow C$. Confounder
  variables have the potential to create a statistical correlation
  between $A$ and $C$, even if there is no direct causal link between
  them. Consequently, confounding variables are factors that can
  obscure or distort the genuine relationship between
  variables. Following the example above, rainy season $\rightarrow$
  $\uparrow$ umbrella sales and rainy season $\rightarrow$ $\uparrow$
  raincoat sales.
\item \emph{Collider variables} are caused by the effect of multiple
  variables: $A \rightarrow B$ and $C \rightarrow B$.  This scenario
  is particularly relevant in chaotic dynamical systems, where most
  variables are affected by multiple causes due to non-linear
  coupling.
\begin{itemize}
  \item A collider variable exhibits \emph{redundant} causes when both
    $A$ and $C$ contribute to the same effect or outcome of $B$,
    creating overlapping or duplicative influences on the outcome.
    Consequently, redundant causes result in multiple pathways to the
    same effect. For instance, both hard work and high intelligence
    can independently contribute to the good grades of a student. Note
    that $A$ and $C$ may not necessarily be independent. 
  \item A collider variable is caused from \emph{synergistic}
    variables if the combined effect of $A$ and $C$ on $B$ surpasses
    their individual effects on $B$ when considered separately. As an
    example, two drugs may be required in tandem to effectively treat
    a condition; when each drug alone is insufficient.
\end{itemize}
\end{itemize}
%

The interactions described above can combine and occur simultaneously,
giving rise to more intricate causal networks. The method proposed
below aims to distinguish between these various interactions, provided
that we have sufficient knowledge of the system.

\section{Method formulation}
\label{sec:method}

\subsection{Fundamentals of information theory}
\label{sec:fundamentals}

Let us introduce the concepts of information theory required to
formulate the problem of IT-causality.  The first question that needs
to be addressed is the meaning of \emph{information}, as it is not
frequently utilized within the fluid dynamics community.  Consider $N$
quantities of interest at time $t$ represented by the vector $\bQ =
[Q_1(t),Q_2(t), \ldots,Q_N(t)]$.  For example, $Q_i(t)$ may be the
velocity or pressure of the flow at a given point in space, temporal
coefficients obtained from proper orthogonal decomposition, or a
spacially-averaged quantity, etc.  We treat $\bQ$ as a random variable
and consider a finite partition of the observable phase space $D
=\{D_1, D_2, \dots, D_{N_Q}\}$, where $N_Q$ is the number of
partitions, such that $D = \cup_{i=1}^{N_Q} D_i$ and $D_i \cap D_j =
\emptyset$ for all $i \neq j$.
We use upper case $Q$ to denote the random variable itself; and lower
case $q$ to denote a particular state or value of $Q$.
The probability of finding the system at state $D_i$ at time $t$ is
$p( \bQ(t) \in D_i )$, that in general depends on the partition $D$.
For simplicity, we refer to the latter probability as $p(\bq)$.

The information contained in the variable $\bQ$ is given
by~\citep{shannon1948}:
\begin{equation}\label{eq:entropy}
  H(\bQ) = \sum_{\bq} -p(\bq) \log_2[p(\bq)]\geq 0,
\end{equation}
where the summation is over all the states (i.e., values) of
$\bQ$. The quantity $H$ is referred to as the Shannon information or
entropy~\citep{shannon1948}. The units of $H$ are set by the base
chosen, in this case `bits' for base 2. For example, consider a fair
coin with $Q\in\{\mathrm{heads},\mathrm{tails}\}$ such that
$p(\mathrm{heads})=p(\mathrm{tails})=0.5$. The information of the
system ``tossing a fair coin $n$ times'' is $H = -\sum 0.5^n
\log_2(0.5^n) = n$ bits, where the summation is carried out across all
possible outcomes (namely, $2^n$).  If the coin is completely biased
towards heads, $p(\mathrm{heads})=1$, then $H = 0$ bits (taking $0\log
0 = 0$), i.e., no information is gained as the outcome was already
known before tossing the coin.  The Shannon information can also be
interpreted in terms of uncertainty: $H(\bQ)$ is the average number of
bits required to unambiguously determine the state $\bQ$.  $H$ is
maximum when all the possible outcomes are equiprobable (indicating a
high level of uncertainty in the system's state) and zero when the
process is completely deterministic (indicating no uncertainty in the
outcome).

The Shannon information of $\bQ$ conditioned on another variable
$\bQ'$ is defined as~\citep{stone2013}:
\begin{equation}
H( \bQ | \bQ' ) = \sum_{\bq,\bq'} -p(\bq,\bq') \log_2[p(\bq|\bq')].
\end{equation}
where $p(\bq|\bq') = p(\bq,\bq')/p(\bq')$ is the conditional
probability distribution, and $p(\bq') = \sum_{\bq} p(\bq,\bq')$ is
the marginal probability distribution of $\bq'$. It is useful to
interpret $H(\bQ|\bQ')$ as the uncertainty in the state $\bQ$ after
conducting the `measurement' of the state $\bQ'$.  If $\bQ$ and $\bQ'$
are independent random variables, then $H(\bQ|\bQ')=H(\bQ)$, i.e.,
knowing the state $\bQ'$ does not reduce the uncertainty in
$\bQ$. Conversely, $H(\bQ|\bQ')=0$ if knowing $\bQ'$ implies that
$\bQ$ is completed determined. Finally, the mutual information between
the random variables $\bQ$ and $\bQ'$ is
\begin{equation}\label{eq:optimal}
   I(\bQ;\bQ') =  H(\bQ) - H(\bQ|\bQ') = H(\bQ') - H(\bQ'|\bQ),
\end{equation}
which is a symmetric measure $I(\bQ;\bQ')=I(\bQ';\bQ)$ representing
the information shared among the variables $\bQ$ and $\bQ'$.  The
mutual information between variables is central to the formalism
presented below. Figure \ref{fig:basics} depicts the relationship
between the Shannon information, conditional Shannon information, and
mutual information.
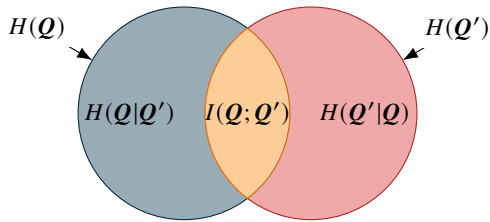
\begin{figure}
\centering
  \begin{tikzpicture}[scale=1.4,>={Latex[length=.2cm]}]
    \colorlet{c1}{myc1}
    \colorlet{c2}{myc2}

    \colorlet{MI}{myc3}
    \pgfmathsetmacro{\dis}{.6}
    \begin{footnotesize}
        \draw[c1!80!black,fill=c1!40] (-\dis,0) circle (1);
        \draw[c2!80!black,fill=c2!40] (\dis,0) circle (1);
        
        \begin{scope}
            \clip (-\dis,0) circle (1);
            \draw[draw=MI,fill=MI!40] (\dis,0) circle (1);
        \end{scope}
        \begin{scope}
            \clip (\dis,0) circle (1);
            \draw[MI!80!black] (-\dis,0) circle (1);
        \end{scope}

        \node[anchor=east] at (-\dis,0) {$H(\bQ |\bQ')$};
        \node[anchor=west] at (+\dis,0) {$H(\bQ'|\bQ)$};

        \node[anchor=center] at (0,0) {$I(\bQ;\bQ')$};

        \draw[<-] (+\dis,0) ++ (30:1)  --+ (30:.6)  node[fill=white,pos=1] {$H(\bQ')$};
        \draw[<-] (-\dis,0) ++ (150:1) --+ (150:.6) node[fill=white,pos=1] {$H(\bQ)$};

    \end{footnotesize}
\end{tikzpicture} 
    \caption{Venn diagram of the Shannon information, conditional
      Shannon information and mutual information between two random
      variables $\bQ$ and $\bQ'$. \label{fig:basics}}
\end{figure}

The definitions above can be extended to continuous random variables
by replacing summation by integration and the probability mass
functions by probability density functions:
\begin{subequations}
  \label{eq:Hc}
  \begin{align}    
    H_c(\bQ) &= \int_{\bq} -\rho(\bq) \log_2[\rho(\bq)] \dd \bq, \\
    H_c(\bQ|\bQ') &= \int_{\bq,\bq'} -\rho(\bq,\bq') \log_2[\rho(\bq|\bq')] \dd \bq\dd \bq', \\
    I_c(\bQ;\bQ') &= H_c(\bQ) - H_c(\bQ|\bQ') = H_c(\bQ') - H_c(\bQ'|\bQ),
  \end{align}
\end{subequations}  
where $H_c$ is referred to as the differential entropy, $\bQ$ and
$\bQ'$ are now continuous random variables, $\rho$ denotes probability
density function, and the integrals are performed over the support set
of $\bQ$ and $\bQ'$.  The differential entropy shares many of the
properties of the discrete entropy. However, it can be infinitely
large, negative, or positive.  The method presented here relies on the
use of mutual information, which is non-negative in the continuous
case. Additionally, it can be shown that if
$\rho(\bq,\bq')\log_2[\rho(\bq,\bq')]$ is Riemann integrable, then
$I(\bQ^\Delta;\bQ'^\Delta) \rightarrow I_c(\bQ;\bQ')$ for $\Delta
\rightarrow 0$, where $\bQ^\Delta$ and $\bQ'^\Delta$ are the quantized
versions of $\bQ$ and $\bQ'$, respectively, defined over a finite
partition with a characteristic size of $\Delta$~\citep{cover2006}. In
the following section, we present our approach using discrete mutual
information; nevertheless, a similar formulation is applicable to the
continuous case.

\subsection{Information-theoretic causality (IT-causality)}
\label{subsec:infocau}

Our objective is to quantify the causality from the components of
$\bQ(t)$ to the future of the variable $Q_j^+ = Q_j(t+\Delta T)$,
where $Q_j$ is one of the components of $\bQ$ and $\Delta T>0$ represents an
arbitrary time lag. Moreover, for each component of $\bQ$, the
causality will be decomposed into \emph{redundant}, \emph{unique}, and
\emph{synergistic} contributions to $Q_j^+$.

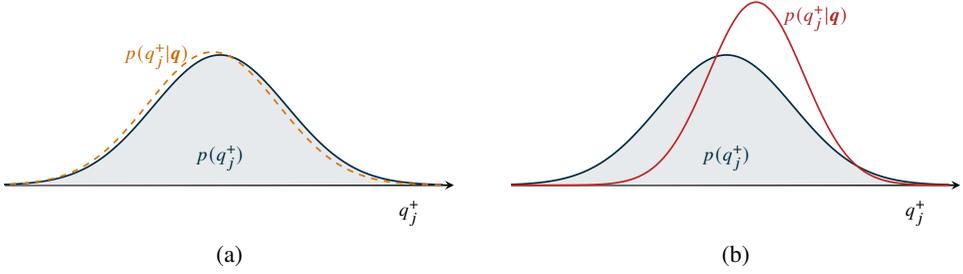
\begin{figure}
  \begin{center}
    \subfloat[]{
\pgfmathdeclarefunction{gauss}{3}{%
  \pgfmathparse{1/(#3*sqrt(2*pi))*exp(-((#1-#2)^2)/(2*#3^2))}%
}
 
\begin{tikzpicture}[scale=.8]

    \def\B{12};   
    \def\S{14};   
    \def\T{11.5}; 

    \def\Bs{4.50};  
    \def\Ss{3.20};  
    \def\Ts{4.40};  

    \def\xmax{25};
    \def\ymin{{-0.15*gauss(\B,\B,\Bs)}};
 
    \begin{axis}[every axis plot post/.append style={
                 mark=none,domain={-0.5*(\xmax)}:{1.08*\xmax},samples=80,smooth},
                 xmin={-0.1*(\xmax)}, xmax=\xmax,
                 ymin=\ymin, ymax={1.3*gauss(\B,\B,\Bs)},
                 axis lines=middle,
                 axis line style=thick,
                 enlargelimits=upper, 
                 ticks=none,
                 xlabel=$q_j^+$,
                 y axis line style={opacity=0},
                 x label style={at={(axis description cs:0.9,0.05)},anchor=north},
                 width=9cm, height=5cm,
                ]

      \addplot[name path=B,thick,black!10!myc1,fill=myc1!10] {gauss(x,\B,\Bs)};
      \addplot[name path=T,thick,black!10!myc3,dashed ] {gauss(x,\T,\Ts)};
 
      \node[above,black!20!myc1 ] at (\B,{gauss(.1*\B,\B,\Bs)}) {$p(q_j^+)$};
      \node[left,black!20!myc3 ] at 
        (.9*\T,{gauss(.9*\T,\T,\Ts)}) {$p(q_j^+| \bq )$};

    \end{axis}

\end{tikzpicture}}
    \hspace{0.5cm}
    \subfloat[]{
\pgfmathdeclarefunction{gauss}{3}{%
  \pgfmathparse{1/(#3*sqrt(2*pi))*exp(-((#1-#2)^2)/(2*#3^2))}%
}
 
\begin{tikzpicture}[scale=.8]

    \def\B{12};   
    \def\S{14};   
    \def\T{11.5}; 

    \def\Bs{4.50};  
    \def\Ss{3.20};  
    \def\Ts{4.40};  

    \def\xmax{25};
    \def\ymin{{-0.15*gauss(\B,\B,\Bs)}};
 
    \begin{axis}[every axis plot post/.append style={
                 mark=none,domain={-0.5*(\xmax)}:{1.08*\xmax},samples=80,smooth},
                 xmin={-0.1*(\xmax)}, xmax=\xmax,
                 ymin=\ymin, ymax={1.3*gauss(\B,\B,\Bs)},
                 axis lines=middle,
                 axis line style=thick,
                 enlargelimits=upper, 
                 ticks=none,
                 xlabel=$q_j^+$,
                 y axis line style={opacity=0},
                 x label style={at={(axis description cs:0.9,0.05)},anchor=north},
                 width=9cm, height=5cm,
                ]

      \addplot[name path=B,thick,black!10!myc1,fill=myc1!10] {gauss(x,\B,\Bs)};
      \addplot[name path=S,thick,black!10!myc2 ] {gauss(x,\S,\Ss)};
 
      \node[above,black!20!myc1 ] at (\B,{gauss(.1*\B,\B,\Bs)}) {$p(q_j^+)$};
      \node[right,black!20!myc2 ] at 
        (1.1*\S,{gauss(1.1*\S,\S,\Ss)}) {$p(q_j^+| \bq )$};

    \end{axis}

\end{tikzpicture}}
  \end{center}
\caption{ Dissimilarity between $p(q_j^+)$ and $p(q_j^+|\bq)$
  contributing to $I(Q_j^+;\bQ)$.  Examples of (a) $p(q_j^+|\bq)$
  resembling $p(q_j^+)$, which barely contributes to $I(Q_j^+;\bQ)$;
  and (b) $p(q_j^+|\bq)$ different from $p(q_j^+)$, which increases
  the value of $I(Q_j^+;\bQ)$. IT-causality from $\bQ$ to $Q_j^+$ is
  quantified by the expectation of
  $\log_2[p(q_j^+|\bq)/p(q_j^+)]$.  \label{fig:interpretation_mutual}}
\end{figure}

The theoretical foundation of the method is rooted in the forward
propagation of information in dynamical systems. Let us consider the
information in the variable $Q_j^+$, given by $H(Q_j^+)$. Assuming
that all the information in $Q_j^+$ is determined by the past states
of the system, we can write the equation for the forward propagation
of information~\citep{Lozano2022}
\begin{equation}
  \label{eq:conservation_info}
 H(Q_j^+) = \Delta I(Q_j^+; \bQ) + \Delta I_{\text{leak}\rightarrow j},
\end{equation}
where $\Delta I(Q_j^+; \bQ)$ is the information flow from $\bQ$ to
$Q_j^+$, and $\Delta I_{\text{leak}\rightarrow j}$ is the information
\emph{leak}, representing the causality from unobserved variables that
influence the dynamics of $Q_j^+$ but are not part of
$\bQ$. The information leak can be expressed in
closed form as a function of the observed variables as
\begin{equation}
  \Delta I_{\text{leak}\rightarrow j} = H(Q_j^+|\bQ),
\end{equation}  
that is the uncertainty in $Q_j^+$ given the information in $\bQ$. The
amount of available information about $Q_j^+$ given $\bQ$ is
\begin{equation}
H(Q_j^+)-\Delta I_{\text{leak}\rightarrow j}= \Delta I(Q_j^+; \bQ)= H(Q_j^+)-H(Q_j^+|\bQ)=I(Q_j^+;\bQ),
\end{equation}
which is the mutual information between $Q_j^+$ and $\bQ$,
\begin{equation}
  \label{eq:mutual_info_decom}
  I(Q_j^+;\bQ) = \sum_{q_j^+,\bq} p(q_j^+,\bq) \log_2\left(
  \frac{p(q_j^+|\bq)}{p(q_j^+)}\right) = \sum_{q_j^+,\bq} p(q_j^+,\bq)
  \log_2\left( \frac{p(q_j^+,\bq)}{p(q_j^+)p(\bq)}\right).
\end{equation}
%
%
Equation (\ref{eq:mutual_info_decom}) quantifies the average
dissimilarity between $p(q_j^+)$ and $p(q_j^+|\bq)$. In terms of the
Kullback-Leibler divergence~\citep{kullback1951}, it measures the
dissimilarity between $p(q_j^+,\bq)$ and the distribution presumed
under the assumption of independence between $Q_j^+$ and $\bQ$,
viz. $p(q_j^+)p(\bq)$. Hence, causality here is assessed by examining
how the probability of $Q_j^+$ changes when accounting for $\bQ$.
Figure~\ref{fig:interpretation_mutual} provides an interpretation of
the quantification of causality based on
Eq.~(\ref{eq:mutual_info_decom}).
%
%

The next step involves decomposing $I(Q_j^+;\bQ)$ into its unique,
redundant, and synergistic components as
\begin{equation}
  \label{eq:conservation_info_2}
I(Q_j^+;\bQ) = \sum_{i=1}^N \Delta I_{i\rightarrow j}^U + \sum_{\bi
  \in \mathcal{C}} \Delta I_{\bi \rightarrow j}^R + \sum_{\bi\in
  \mathcal{C}} \Delta I_{\bi \rightarrow j}^S,
\end{equation}
where $\Delta I_{i\rightarrow j}^U$ is the unique causality from $Q_i$
to $Q_j^+$, $\Delta I_{\bi \rightarrow j}^R$ is the redundant
causality among the variables in $\bQ_{\bi}$ with $\bi = [i_1, i_2,
  \ldots]$ being a collection of indices, $\Delta I_{\bi \rightarrow j}^S$
is the synergistic causality from the variables in $\bQ_{\bi}$, and
$\mathcal{C}$ is the set of all the combinations taken from 1 to $N$
with more than one element and less than or equal to $N$ elements.
For $N=4$, Eq. (\ref{eq:conservation_info_2}) can be expanded as
\begin{subequations}
\begin{align}    
  I(Q_j^+;\bQ) &=  \Delta I_{1\rightarrow j}^U +  \Delta I_{2\rightarrow j}^U  +  \Delta I_{3\rightarrow j}^U +  \Delta I_{4\rightarrow j}^U \\  
  &+  \Delta I_{12 \rightarrow j}^R + \Delta I_{13 \rightarrow j}^R + \Delta I_{14 \rightarrow j}^R  +
  \Delta I_{23 \rightarrow j}^R + \Delta I_{24 \rightarrow j}^R + \Delta I_{34 \rightarrow j}^R  + \\
  &+  \Delta I_{12 \rightarrow j}^S + \Delta I_{13 \rightarrow j}^S + \Delta I_{14 \rightarrow j}^S  +
  \Delta I_{23 \rightarrow j}^S + \Delta I_{24 \rightarrow j}^S + \Delta I_{34 \rightarrow j}^S  + \\  
  &+ \Delta I_{123 \rightarrow j}^R + \Delta I_{124 \rightarrow j}^R + 
  \Delta I_{134 \rightarrow j}^R + \Delta I_{234 \rightarrow j}^R + \\
    &+ \Delta I_{123 \rightarrow j}^S + \Delta I_{124 \rightarrow j}^S + 
  \Delta I_{134 \rightarrow j}^S + \Delta I_{234 \rightarrow j}^S \\
  &+ \Delta I_{1234 \rightarrow j}^R + \Delta I_{1234 \rightarrow j}^S.
\end{align}    
\end{subequations}

The source of causality might change depending on the value of
$Q_j^+$. For example, $Q_1$ can be only causal to positive values of
$Q_j^+$, whereas $Q_2$ can be only causal to negative values of
$Q_j^+$. For that reason, we define the specific mutual
information~\citep{DeWeese1999} from $\bQ_{\bi}$ to a particular event
$Q_j^+=q_j^+$ as
\begin{equation}
  \label{eq:specific_mutual_2}
   \Is(Q_j^+ = q_j^+;\bQ_{\bi}) = \sum_{\bq_{\bi}} p(\bq_{\bi} | q_j^+)
  \log_2\left( \frac{p(q_j^+|\bq_{\bi})}{p(q_j^+)} \right) \geq 0.
\end{equation}
Note that the specific mutual information is a function of the random
variable $\bQ_{\bi}$ (which encompasses all its states) but only a
function of one particular state of the target variable (namely,
$q_j^+$). For the sake of simplicity, we will use $\Is_{\bi}(q_j^+) =
\Is(Q_j^+ = q_j^+;\bQ_{\bi})$.
Similarly to Eq. (\ref{eq:mutual_info_decom}), the specific mutual
information quantifies the dissimilarity between $p(q_j^+)$ and
$p(q_j^+|\bq)$ but in this case for the particular state
$Q_j^+=q_j^+$. The mutual information between $Q_j^+$and $\bQ_{\bi}$
is recovered by $I(Q_j^+;\bQ_{\bi}) = \sum_{q_j^+} p(q_j^+)
\Is_{\bi}(q_j^+)$.

\begin{figure}
    \centering
    \begin{tikzpicture}
        \node[anchor=north west] (fig) at (0,0) {\includegraphics[width=.8\tw]{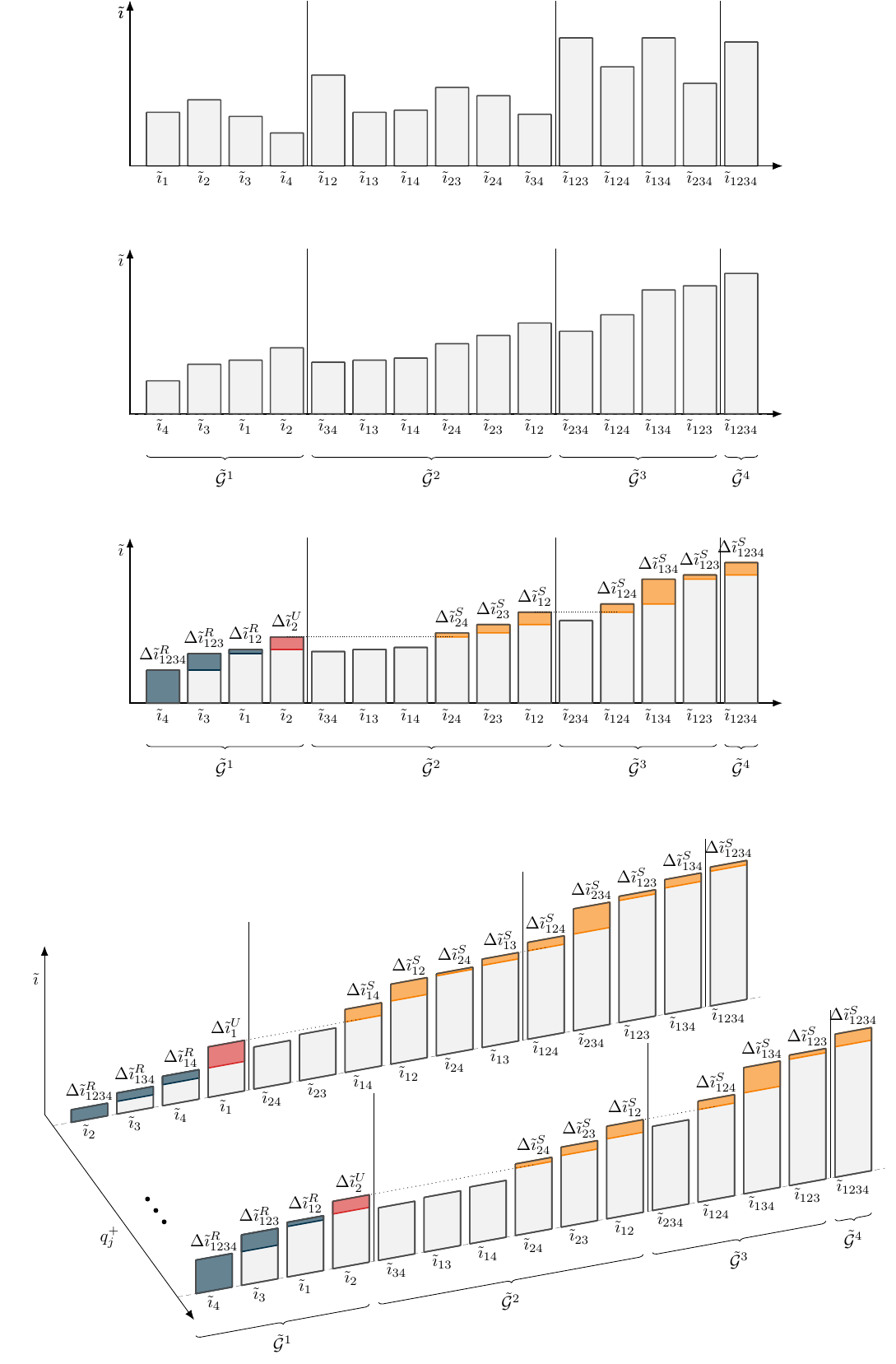}};
        \node at (0,0) {(a)};
        \node at (0,-3.5) {(b)};
        \node at (0,-7.5) {(c)};
        \node at (0,-12) {(d)};
    \end{tikzpicture}
\caption{ Schematic of the steps involved in the calculation of
  specific causalities. For a given state $Q_j^+=q_j^+$, the panels
  illustrate: (a) all possible specific mutual information values for
  a collection of four variables; (b) tuples of specific mutual
  information with the components organized in ascending order; (c)
  the increments corresponding to specific redundant (blue), unique
  (red), and synergistic (yellow) causalities; and (d) examples of
  specific causalities for different states of
  $Q_j^+$. \label{fig:method}}
\end{figure}
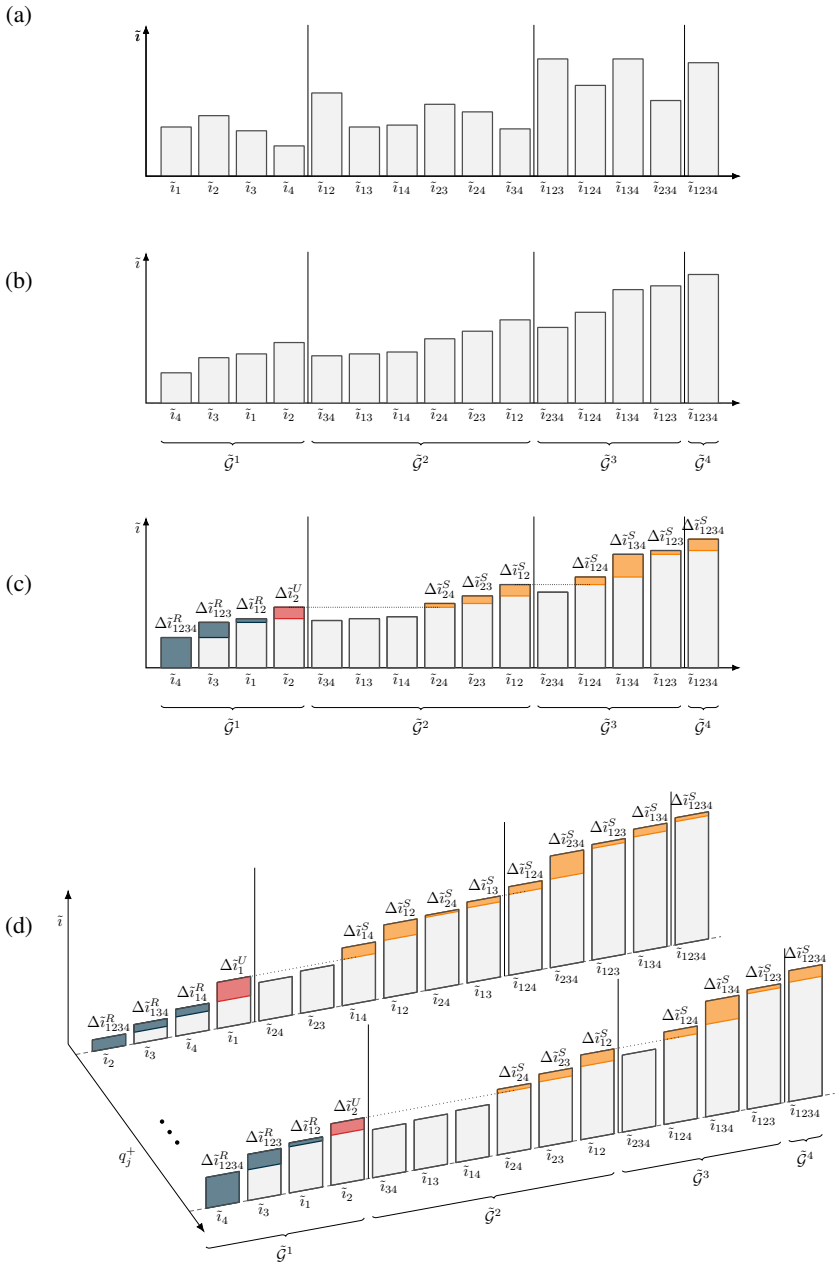

We are now in the position of outlining the steps involved in the
calculation of redundant, unique, and synergistic causalities
(figure~\ref{fig:method}). The particular choices of the method were
made to comply with the intuition behind mediator, confounder, and
collider interactions (\S\ref{sec:basic_causal}) along with the ease
of interpretability of the results.  The reader is referred to
Appendix~\ref{sec:appendixA} for the formal definitions.  For a given
value $Q_j^+=q_j^+$, the \emph{specific} redundant, unique, and
synergistic causalities are calculated as follows:
\begin{enumerate}
\item The specific mutual information are computed for all possible
  combinations of variables in $\bQ$. This includes specific mutual
  information of order one ($\Is_1, \Is_2, \ldots$), order two
  ($\Is_{12}, \Is_{13}, \ldots$), order three ($\Is_{123}, \Is_{124},
  \ldots$), and so forth. One example is shown in
  figure~\ref{fig:method}(a).
\item The tuples containing the specific mutual information of order
  $M$, denoted by $\GIs^M$, are constructed for $M=1,\ldots,N$.  The
  components of each $\GIs^M$ are organized in ascending order as shown in
  figure~\ref{fig:method}(b). 
\item The specific redundant causalities, $\Delta \Is^R_{\bi}(q_j^+)$,
  are defined as the increments of information in $\GIs^1$ common to
  all the components of $\bQ_{\bi}$ (blue contributions in
  figure~\ref{fig:method}c).
\item The specific unique causality, $\Delta \Is^U_{i}(q_j^+)$, is
  defined as the increment of information from $Q_i$ that cannot be
  obtained from any other individual variable $Q_k$ with $k\neq i$
  (red contribution in figure~\ref{fig:method}c).
\item The specific synergistic causalities, $\Delta
  \Is^S_{\bi}(q_j^+)$, are defined as the increments of information
  due to the joint effect of the components of $\bQ_{\bi}$ in $\GIs^M$
  with $M>1$ (yellow contributions in figure~\ref{fig:method}c). The
  first increment is computed using as reference the largest specific
  mutual information from the previous tuple (dotted line in
  figure~\ref{fig:method}c).
\item The specific redundant, unique and synergistic causalities that
  do not appear in the steps above are set to zero.
\item The steps (i) to (vi) are repeated for all the states of $Q_j^+$
  (figure~\ref{fig:method}d).
\item The IT-causalities (redundant, unique, and synergistic) are
  obtained as the expectation of their corresponding specific values
  with respect to $Q_j^+$,
  \begin{subequations}
    \begin{align}  
      \Delta I^R_{\bi \rightarrow j} &= \sum_{q_j^+} p(q_j^+) \Delta \Is^R_{\bi}(q_j^+), \\
      \Delta I^U_{i\rightarrow j}   &= \sum_{q_j^+} p(q_j^+) \Delta \Is^U_{i}(q_j^+), \\
      \Delta I^S_{\bi\rightarrow j} &= \sum_{q_j^+} p(q_j^+) \Delta \Is^S_{\bi}(q_j^+). 
    \end{align}  
  \end{subequations}
\item 
  Finally, we define the average order of the specific IT-causalities
  with respect to $Q_j^+$ as
  \begin{equation}
    N^{\alpha}_{\bi \rightarrow j} = \sum_{q_j^+} p(q_j^+)
    n^{\alpha}_{\bi \rightarrow j}(q_j^+),
  \end{equation}
  where $\alpha$ denotes R, U, or S, $n^{\alpha}_{\bi \rightarrow
    j}(q_j^+)$ is the order of appearance of $\Delta
  \Is^{\alpha}_{\bi}(q_j^+)$ from left to right as in the example
  shown in figure~\ref{fig:method}.  The values of $N^{\alpha}_{\bi
    \rightarrow j}$ will be used to plot $\Delta I^{\alpha}_{\bi
    \rightarrow j}$ following the expected order of appearance of
  $\Delta \Is^{\alpha}_{\bi \rightarrow j}$.
\end{enumerate}
%
\subsection{Simple example of IT-causality}
\label{subsec:simple}

\begin{figure}
  \begin{center}
    \includegraphics[width=1.05\textwidth]{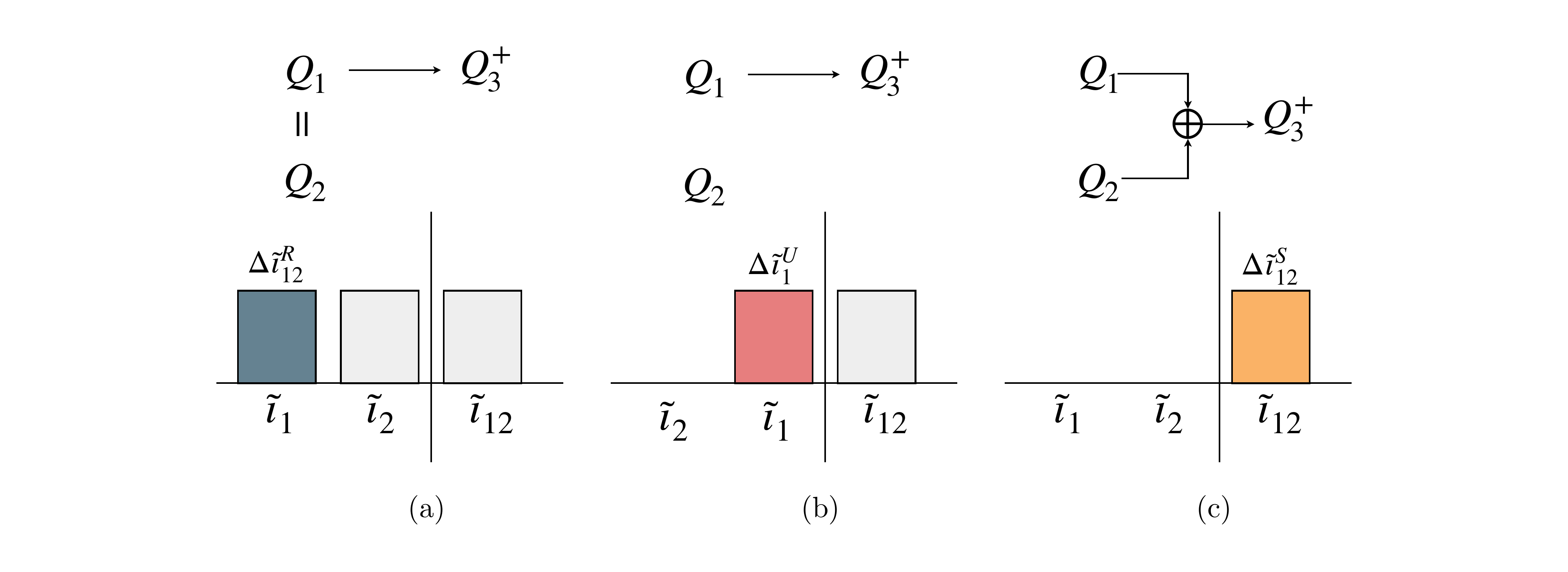}
 \end{center}
\caption{ Schematic of simple examples (top panels) and associated
  specific mutual information (bottom panels) for (a) duplicated input
  (pure redundant causality), (b) output equal to first input (pure
  unique causality), and (c) exclusive-OR output (pure synergistic
  causality). The schematics of the specific mutual information
  apply to both states $Q_3^+=0$ and $Q_3^+=1$. \label{fig:simple}}
\end{figure}

We illustrate the concept of redundant, unique, and synergistic
causality in three simple examples. The examples represent a system
with two inputs $Q_1$ and $Q_2$ and one output $Q_3^+ = f(Q_1,Q_2)$.
The inputs can take two values, $\{0, 1\}$, randomly and independently
distributed, each with a probability of 0.5.  The causal description
of the system is characterized by the four components:
\begin{equation}
  \label{eq:simple}
  H(Q_3^+) = \Delta I^U_{1 \rightarrow 3} + \Delta I^U_{2 \rightarrow 3} +
  \Delta I^R_{12 \rightarrow 3} + \Delta I^S_{12 \rightarrow 3},
\end{equation}  
where $\Delta I_{\text{leak} \rightarrow 3} = 0$ as $H(Q_3^+ | Q_1,
Q_2) =0$.  The results for the three cases are summarized in in
figure~\ref{fig:simple}.

The first example represents a system in which $Q_2=Q_1$ (duplicated
input) and the output is given by $Q_3^+=Q_1$. In this case, both
$Q_1$ and $Q_2$ provide the same information about the output and the
only non-zero term in Eq. (\ref{eq:simple}) is the redundant causality
$\Delta I^R_{12 \rightarrow 3} = 1$ bit.  In the second example, the
output is given by $Q_3^+ = Q_1$ with no dependence on $Q_2$, which
only results in the unique causality $\Delta I^U_{1\rightarrow 3} = 1$
bit.  In the last example, the output is given by the exclusive-OR
operator: $Q_3^+=Q_1 \oplus Q_2$ such that $Q_3^+ = 1$ if $Q_1\neq
Q_2$ and $Q_3^+ = 0$ otherwise. In this case, the output behaves
randomly when observing $Q_1$ or $Q_2$ independently.  However, the
outcome is completely determined when the joint variable $[Q_1,Q_2]$
is considered. Hence, $[Q_1,Q_2]$ contains more information than their
individual components and all the causality comes from the synergistic
causality $\Delta I^S_{12 \rightarrow 3} = 1$ bit.
%


\subsection{Contribution of different intensities to IT-causality}
\label{subsec:intensities}

The IT-causalities can be decomposed in terms of different
contributions from $\bQ_{\bi}$ and $Q_j^+$. For the case of the unique
causality, the IT-causality as function of $q_i$ and $q_j^+$ is
denoted by $\Delta \Iss^U_{i \rightarrow j}$ and it is such that
\begin{equation}
  \label{eq:cau_per_int}
  \Delta I^U_{i \rightarrow j} = \sum_{q_i} \sum_{q^+_j} \Delta \Iss^U_{i \rightarrow j}(q_i,q_j^+).
\end{equation}
The expression for $\Delta \Iss^U_{i \rightarrow j}$ is obtained
inverting Eq.~(\ref{eq:cau_per_int})
\begin{equation}
  \Delta \Iss^U_{i \rightarrow j} = 
  \begin{cases}
  p(q^+_j) p( q_i | q^+_j ) \log_2\left( \frac{p(q^+_j | q_i )}{p(q^+_j)} \right) - p(q^+_j) \frac{\Is_l}{N_Q}, & \text{for} \ \Is_i\geq \Is_l, \\
   0,              & \text{otherwise},
\end{cases}
\end{equation}
where $\Is_l$ is the second largest member in $\GIs^1$ and $N_Q$ is the
total number of states of $q_i$.

Analogous definitions can be written for the redundant and synergistic
causalities. Note that $\Delta \Iss^U_{i \rightarrow j}$ may be
negative for a given $q_i$--$q_j^+$ pair; although the sum of all the
components is non-negative. Positive values of $\Delta \Iss^U_{i
  \rightarrow j}$ are \emph{informative} (the pair $q_i$--$q_j^+$
occurs more frequently than would be expected) and negative values are
\emph{misinformative} (the pair $q_i$--$q_j^+$ occurs less frequently
than would be expected).

\subsection{Properties of IT-causality}
\label{subsec:properties}

We discuss some properties of the IT-causality.
\begin{itemize}
\item \emph{Non-negativity}. All the terms in
  Eq. (\ref{eq:conservation_info_2}) are non-negative by the
  definition of the redundant, unique and synergistic causalities, and
  the non-negativity of the specific mutual
  information~\citep{DeWeese1999}.
  %
\item \emph{Reconstruction of individual mutual information}. The
  mutual information between $Q_i$ and $Q_j^+$ is equal to the unique
  and redundant causalities containing $Q_i$
  \begin{equation}\label{eq:individual_causality}
    I(Q_i; Q_j^+ ) = \Delta I_{i\rightarrow j}^U +
    \sum_{\boldsymbol{i}\in \mathcal{C}_i} \Delta I_{\boldsymbol{i}
      \rightarrow j}^R,
  \end{equation}
  where $\mathcal{C}_i$ is the set of the combinations in
  $\mathcal{C}$ containing the variable $Q_i$. This condition aligns
  with the notion that the information shared between $Q_i$ and
  $Q_j^+$ comprises contributions from unique and redundant
  information. However, there is no contribution from synergistic
  information, as the latter only arises through the combined effects
  of variables. This property, along with the non-negativity and
  forward propagation of information, enables the construction of the
  causality diagrams as depicted in figure~\ref{fig:diagram_mutuals}
  for two and three variables.
  %
\begin{figure}
  \begin{center}
    \subfloat[]{\begin{tikzpicture}[scale=1.,thin,>={Latex[length=.1cm]}]
    \colorlet{c2}{myc2}
    \colorlet{c1}{myc3}

    \colorlet{MI}{myc1}
    \pgfmathsetmacro{\dis}{.6}
    \begin{scriptsize}

        \draw[->] (-\dis,0) ++ (130:1)  --+ (130:1.)  node[anchor=south,pos=1] {$I(Q^+_j;Q_1,Q_2)$};

        \draw[fill=c1!45] (0,0) ellipse (2.2 and 1.4);

        \draw[fill=c2!45] (-\dis,0) circle (1);
        \draw[fill=c2!45] (\dis,0) circle (1);
        
        \begin{scope}
            \clip (-\dis,0) circle (1);
            \draw[fill=MI!45] (\dis,0) circle (1);
        \end{scope}
        \begin{scope}
            \clip (\dis,0) circle (1);
            \draw (-\dis,0) circle (1);
        \end{scope}

        \node[anchor=east] at (-\dis,0) {$\Delta I_{1\to j}^U$};
        \node[anchor=west] at (+\dis,0) {$\Delta I_{2\to j}^U$};
        \node              at (    0,0) {$\Delta I_{12\to j}^R$};
        \node              at ( 0, 1.2) {$\Delta I_{12\to j}^S$};


        \draw[->] (+\dis,0) ++ (-40:1)  --+ (-40:1.)  node[anchor=north,pos=1] {$I(Q^+_j;Q_2)$};
        \draw[->] (-\dis,0) ++ (220:1)  --+ (220:1.)  node[anchor=north,pos=1] {$I(Q^+_j;Q_1)$};

    \end{scriptsize}
\end{tikzpicture}}
    \subfloat[]{

\colorlet{c1}{myc1}
\colorlet{c2}{myc2}
\colorlet{c3}{myc3}

\colorlet{c1s}{c1!50}
\colorlet{c2s}{c2!50}
\colorlet{c3s}{c3!50}

\begin{tikzpicture}[scale=1.25,thin,>={Latex[length=.1cm]}
]
    
    \begin{scriptsize}
    \pgfmathsetmacro{\rdi}{1}
    \pgfmathsetmacro{\rdiO}{.9*\rdi}
    \pgfmathsetmacro{\dis}{.7}

    \coordinate (O3) at (90:\dis);
    \coordinate (O2) at (-30:\dis);
    \coordinate (O1) at (-150:\dis);

    \coordinate (O4) at (270:1.2*\dis);
    \coordinate (O5) at ( 30:1.2*\dis);
    \coordinate (O6) at (150:1.2*\dis);
   
    \draw[thin,->] (0,0) --++ (120:2.4) node[anchor=south]{$I(Q_j^+;Q_1,Q_2,Q_1)$};
    \draw[fill=c3!45] (0,.05) ellipse (2.1 and 2.05 );

    \draw[fill=c3!45] (O4) circle (1);
    \draw[fill=c3!45] (O5) circle (1);
    \draw[fill=c3!45] (O6) circle (1);


    \draw[fill=c2!45] (O1) circle (1);
    \draw[fill=c2!45] (O2) circle (1);
    \draw[fill=c2!45] (O3) circle (1);
    
    \begin{scope}
        \clip (O2) circle (1);
        \draw[fill=c1!45] (O1) circle (1);
    \end{scope}

    \begin{scope}
        \clip (O2) circle (1);
        \draw[fill=c1!45] (O3) circle (1);
    \end{scope}

    \begin{scope}
        \clip (O3) circle (1);
        \draw[fill=c1!45] (O1) circle (1);
    \end{scope}

    \draw[name path=C1] (O1) circle (1);
    \draw[name path=C2] (O2) circle (1);
    \draw[name path=C3] (O3) circle (1);



    \draw[thin,->] (O2)+(-55:\rdi) --++ (-55:2) node[anchor=north]{$I(Q_j^+;Q_2)$};
    \draw[thin,->] (O1)+(235:\rdi) --++ (235:2) node[anchor=north]{$I(Q_j^+;Q_1)$};
    \draw[thin,->] (O3)+( 50:\rdi) --++ ( 50:2) node[anchor=south]{$I(Q_j^+;Q_3)$};

   
    \node[scale=.9] at (90:1*\rdi)   {$\Delta I^U_{3\to j}$};
    \node[scale=.9] at (-30:1*\rdi)  {$\Delta I^U_{2\to j}$};
    \node[scale=.9] at (-150:1*\rdi) {$\Delta I^U_{1\to j}$};

    \node[scale=.9] at ( 30:.6*\rdi) {$\Delta I^R_{23\to j}$};
    \node[scale=.9] at (150:.6*\rdi) {$\Delta I^R_{13\to j}$};
    \node[scale=.9] at (270:.6*\rdi) {$\Delta I^R_{12\to j}$};

    \node[scale=.9] at (0:0) {$\Delta I^R_{123\to j}$};

    \node[scale=.9] at ( 30:1.55*\rdi) {$\Delta I^S_{23\to j}$};
    \node[scale=.9] at (150:1.55*\rdi) {$\Delta I^S_{13\to j}$};
    \node[scale=.9] at (270:1.55*\rdi) {$\Delta I^S_{12\to j}$};

    \node at (90:1.9*\rdi) {$\Delta I^S_{123\to j}$};


    \end{scriptsize}
\end{tikzpicture}}
 \end{center}
\caption{Diagram of the decomposition into redundant, unique, and
  synergistic causalities and contributions to total and individual
  mutual information for (a) two variables and (b) three
  variables.\label{fig:diagram_mutuals}}
\end{figure}
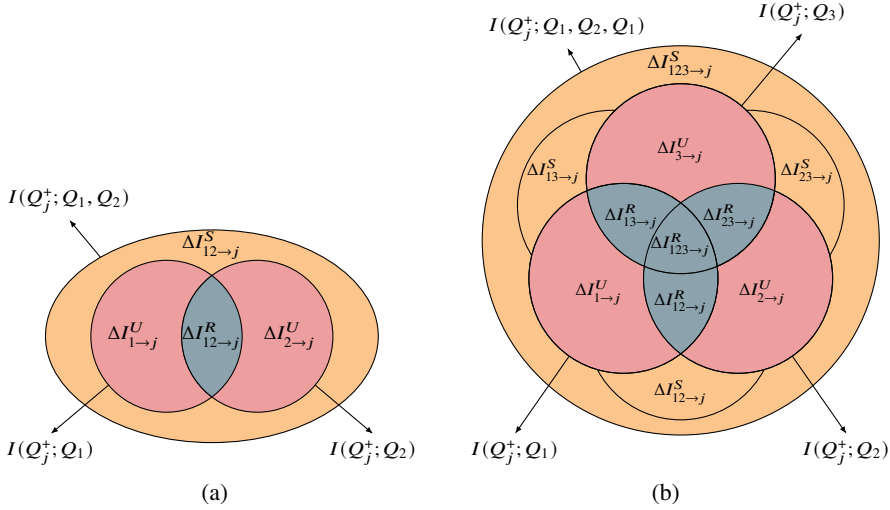
  \item \emph{Zero-causality property}. If $Q_j^+$ is independent of
    $Q_i$, then $\Delta I^R_{\bi \rightarrow j} = 0$ for $\bi \in
    \mathcal{C}_i$ and $\Delta I^U_{i \rightarrow j}=0$ as long as
    $Q_i$ is observable.
\item \emph{Invariance under invertible transformations}. The
  redundant, unique, and synergistic causalities are invariant under
  invertible transformations of $\bQ$. This property follows from the
  invariance of the mutual information~\citep{cover2006}.
%
\end{itemize}

\section{Validation in stochastic systems}
\label{sec:validation}

We discuss four illustrative examples highlighting key distinctions
between IT-causality and time cross-correlations. These examples are
not representative of any specific dynamical system. However, the
phenomena they portray are expected to emerge to varying degrees in
more complex systems. For comparisons, the ``causality'' based on the
time cross-correlation from $Q_i$ to $Q_j$ is defined as
\begin{equation}
\label{eq:Cij}
C_{i\rightarrow j} = \frac{ |\sum_{n=1}^{N_t} Q_i'(t_n)  Q_j'(t_n+\Delta T) | }
{\left( \sum_{n=1}^{N_t} Q_i'^2(t_n) \right)^{1/2}
\left( \sum_{n=1}^{N_t} Q_j'^2(t_n) \right)^{1/2}} \geq 0,
\end{equation}
where $Q_i'(t_n)$ signifies the fluctuating component of $Q_i(t_n)$ at
time $t_n$ with respect to its mean value, and $N_t$ is the total
number of time steps considered for the analysis.  The values of
Eq.~(\ref{eq:Cij}) are bounded between 0 and 1.

In all the scenarios discussed below, we consider a system with three
variables $Q_1(t_n)$, $Q_2(t_n)$, and $Q_3(t_n)$ with discrete times $t_n = n$. 
The systems is initialized at the first step with $Q_1(1) = Q_2(1) = Q_3(1)
= 0$. A time-varying stochastic forcing, denoted by $W_i(t_n)$, acts
on $Q_i(t_n)$ following a Gaussian distribution with a mean of zero
and standard deviation of one. IT-causality is computed for the time
lag $\Delta T=1$ using a partition of 50 uniform bins per
variable. The integration of the systems is carried out over $10^8$
time steps and the first 10,000 steps are excluded from the analysis
to avoid transient effects.
\begin{itemize}
\item Mediator variable ($Q_3 \rightarrow Q_2 \rightarrow Q_1$).
 The first example corresponds to the system:  
\begin{subequations}
  \label{eq:appen1}
  \begin{align}
    Q_1(n+1) &= \sin[Q_2(n)] + 0.01W_1(n), \\
    Q_2(n+1) &= \cos[Q_3(n)] + 0.01W_2(n), \\
    Q_3(n+1) &= 0.9Q_3(n) + 0.1W_3(n),
  \end{align}
\end{subequations}
where $Q_2$ is the mediator variable between $Q_1$ and $Q_3$.  The
results are shown in figure~\ref{fig:appen1}, which includes an
schematic of the functional dependence among variables, the
IT-causality, and time cross-correlations. IT-causality reveals the
prevalence of the unique contributions $\Delta I^U_{3 \rightarrow 3}$,
$\Delta I^U_{3 \rightarrow 2}$, and $\Delta I^U_{2 \rightarrow 1}$, in
addition to some redundant contributions consistent with
figure~\ref{fig:appen1}(a). The unique causalities are compatible with
the functional dependency $Q_3 \rightarrow Q_2 \rightarrow Q_1$. This
is less evident when using correlations, as $C_{1 \rightarrow 1}$,
$C_{2 \rightarrow 1}$, $C_{1 \rightarrow 2}$, and $C_{2 \rightarrow
  2}$ are large with values between 0.5 and 0.95.  IT-causality also
provides information about the amount of information leak
(i.e. missing causality), which is above 50\% for all three variables
due to the effect of the stochastic forcing terms (which assumed to be
unknown). The latter cannot be quantified using correlations.
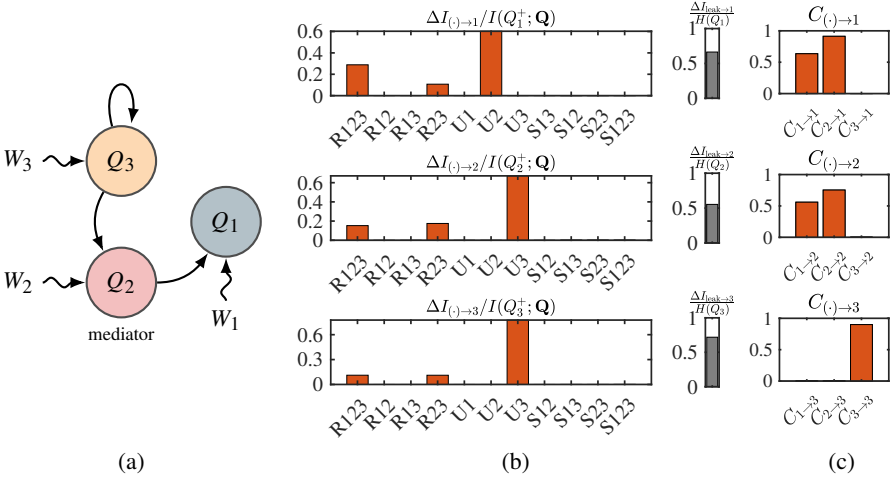
\begin{figure}
  \begin{center}
     \subfloat[]{\raisebox{1.2cm}{\begin{tikzpicture}[cir/.style={circle,draw=black!70!white,,thick,inner sep=.5em},
    >={Latex[length=.2cm]},scale=.8]
    
    \colorlet{cA}{myc2!30!white}
    \colorlet{cB}{myc1!30!white}
    \colorlet{cC}{myc3!30!white}

    \node [cir,fill=cB] (q1) at (.73,0)   {$Q_1$}; 
    \node [cir,fill=cA] (q2) at (-1,-1) {$Q_2$}; 
    \node [cir,fill=cC] (q3) at (-1,+1) {$Q_3$}; 
    
    \node [anchor=north] at (q2.south) {\scriptsize mediator};

    \draw[decorate,decoration={snake,post length=4pt,amplitude=.2em}, thick,->]
    (-2.3,1) node[anchor=east] {$W_3$} -- (q3);
    \draw[decorate,decoration={snake,post length=4pt,amplitude=.2em}, thick,->]
    (-2.3,-1) node[anchor=east] {$W_2$} -- (q2);
    \draw[decorate,decoration={snake,post length=4pt,amplitude=.2em}, thick,->]
    (.73,-1.3) node[anchor=north] {$W_1$} -- (q1);

    \path[thick,->]
        (q3) edge[bend right] node [left] {} (q2)
        (q2) edge[bend right] node [left] {} (q1);
    \path[thick,<-] (q3) edge[loop above] node {} (q3);
    
\end{tikzpicture}}}
    \hspace{0.3cm}
      \subfloat[]{\includegraphics[width=0.43\textwidth]{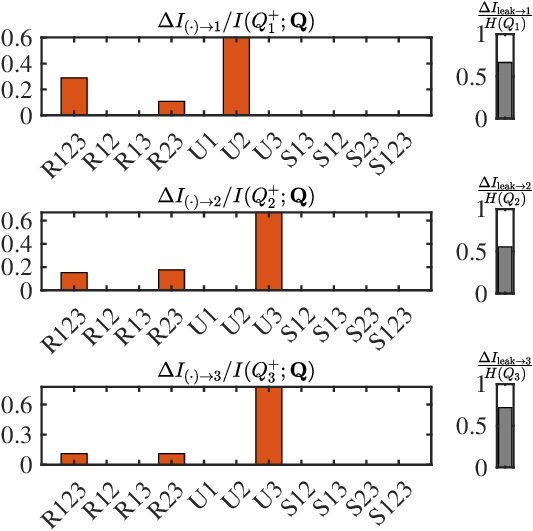}}
      \hspace{0.1cm}
      \subfloat[]{\raisebox{0.15cm}{\includegraphics[width=0.138\textwidth]{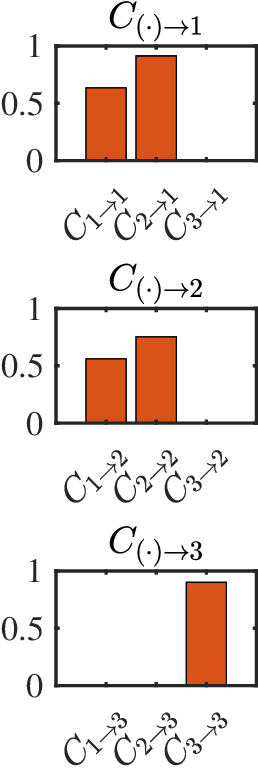}}}
\end{center}
\caption{ System with mediator variable from
  Eq.~(\ref{eq:appen1}). (a) Schematic of the functional dependence
  among variables. $W_i$ represents stochastic forcing to the
  variable.  (b) Redundant, unique and synergistic causalities. The
  grey bar is the information leak. The IT-causalities are ordered from
  left to right according to $N_{\bi \rightarrow
    j}^{\alpha}$. (c) Time cross-correlation between
  variables.  \label{fig:appen1}}
\end{figure}
%
\item Confounder variable ($Q_3 \rightarrow Q_1$ and $Q_3 \rightarrow Q_2$).
 The second example considered is:  
  \begin{subequations}
  \label{eq:appen2}
  \begin{align}
    Q_1(n+1) &= \sin[Q_1(n) + Q_3(n)] + 0.01W_1(n), \label{eq:appen2_a}\\
    Q_2(n+1) &= \cos[Q_2(n) - Q_3(n)] + 0.01W_2(n),\label{eq:appen2_b} \\
    Q_3(n+1) &= 0.9Q_3(n) + 0.1W_3(n),
  \end{align}
\end{subequations}
  where $Q_3$ is a confounder variable to $Q_1$ and $Q_2$.  The
  results are depicted in figure~\ref{fig:appen2}. The influence of
  the confounder variable becomes apparent through the synergistic
  causalities, namely, $\Delta I^S_{13 \rightarrow 1}$ and $\Delta
  I^S_{23 \rightarrow 2}$, as $Q_3$ co-occurs with $Q_1$ and $Q_2$ in
  Eq.~(\ref{eq:appen2_a}) and Eq.~(\ref{eq:appen2_b}),
  respectively. The unique causality $\Delta I^U_{3 \rightarrow 3}$
  dominates $Q_3$. The non-zero redundant causality $\Delta I^R_{123
    \rightarrow 3}$ implies that all variables contain information
  about the future of $Q_3$, but that information is already contained
  in the past of $Q_3$. When considering correlations, drawing robust
  conclusions regarding the interplay among variables presents a
  greater challenge due to the strong correlations observed across all
  possible pairs of variables, with values ranging between 0.6 and 1.
\begin{figure}
\begin{center}
     \subfloat[]{\raisebox{1.2cm}{\begin{tikzpicture}[cir/.style={circle,draw=black!70!white,,thick,inner sep=.5em},
    >={Latex[length=.2cm]},scale=.8]
    
    \colorlet{cA}{myc2!30!white}
    \colorlet{cB}{myc1!30!white}
    \colorlet{cC}{myc3!30!white}

    \node [cir,fill=cB] (q1) at (.73,0) {$Q_1$}; 
    \node [cir,fill=cA] (q2) at (-1,-1) {$Q_2$}; 
    \node [cir,fill=cC] (q3) at (-1,+1) {$Q_3$}; 

    \node [anchor=west] at (q3.north) {\quad \scriptsize confounder};

    \draw[decorate,decoration={snake,post length=4pt,amplitude=.2em}, thick,->]
    (-2.3,1) node[anchor=east] {$W_3$} -- (q3);
    \draw[decorate,decoration={snake,post length=4pt,amplitude=.2em}, thick,->]
    (-2.3,-1) node[anchor=east] {$W_2$} -- (q2);
    \draw[decorate,decoration={snake,post length=4pt,amplitude=.2em}, thick,->]
    (.73,-1.3) node[anchor=north] {$W_1$} -- (q1);

    \path[thick,->]
        (q3) edge[bend left] node [right] {} (q1)
        (q3) edge[bend right] node [left] {} (q2);
    \path[thick,<-] (q1) edge[loop right] node {} (q1);
    \path[thick,<-] (q2) edge[loop below] node {} (q2);
    \path[thick,<-] (q3) edge[loop above] node {} (q3);

\end{tikzpicture}}}
    \hspace{0.2cm}
      \subfloat[]{\includegraphics[width=0.43\textwidth]{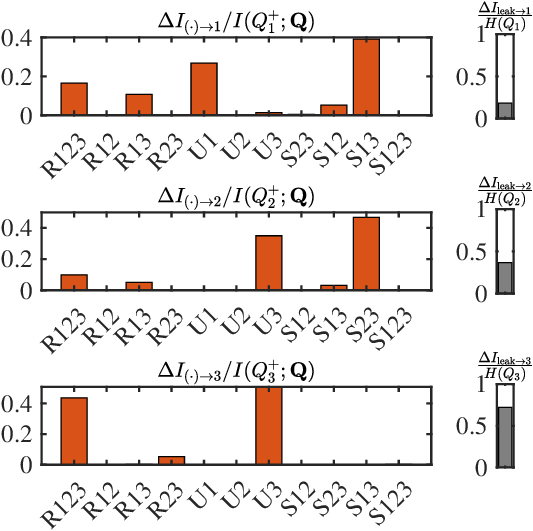}}
      \hspace{0.1cm}
      \subfloat[]{\raisebox{0.15cm}{\includegraphics[width=0.138\textwidth]{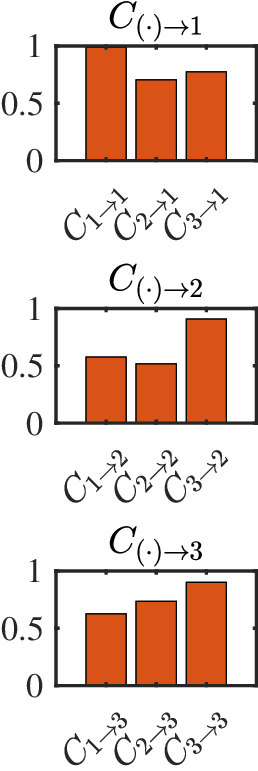}}}
\end{center}
\caption{ System with a confounder variable from
  Eq.~(\ref{eq:appen2}). (a) Schematic of the functional dependence
  among variables. $W_i$ represents stochastic forcing to the
  variable.  (b) Redundant, unique and synergistic causalities. The
  grey bar is the information leak. The IT-causalities are ordered from
  left to right according to $N_{\bi \rightarrow
    j}^{\alpha}$. (c) Time cross-correlation between
  variables.  \label{fig:appen2}}
\end{figure}
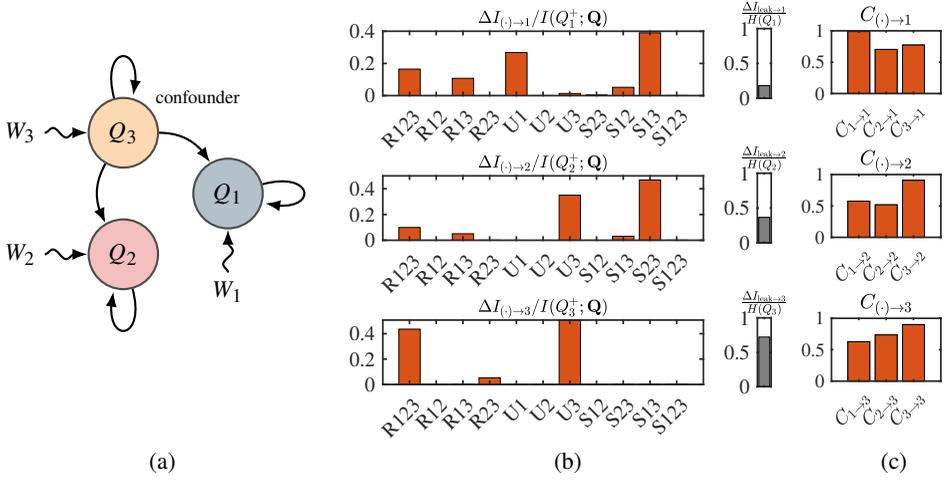
%
\item Collider with synergistic variables ($ [Q_2, Q_3] \rightarrow
  Q_1$).  The third example correspond to the system:
  \begin{subequations}
  \label{eq:appen3}
  \begin{align}
    Q_1(n+1) &= \sin[Q_2(n)Q_3(n)] + 0.001W_1(n), \\
    Q_2(n+1) &= 0.9Q_2(n) + 0.1W_2(n), \\
    Q_3(n+1) &= 0.9Q_3(n) + 0.1W_3(n),
  \end{align}
\end{subequations}
  where $Q_2$ and $Q_3$ work synergistically to influence
  $Q_1$. Essentially, $Q_2 Q_3$ behaves as a single random variable
  acting on $Q_1$. The results, depicted in figure~\ref{fig:appen3},
  clearly reveal the synergistic effect of $Q_2$ and $Q_3$ on $Q_1$,
  as evidenced by $\Delta I^S_{23 \rightarrow 1}$.
  It is worth noting that, in this case, correlations do not hint at
  any influence of $Q_2$ and $Q_3$ on $Q_1$.
  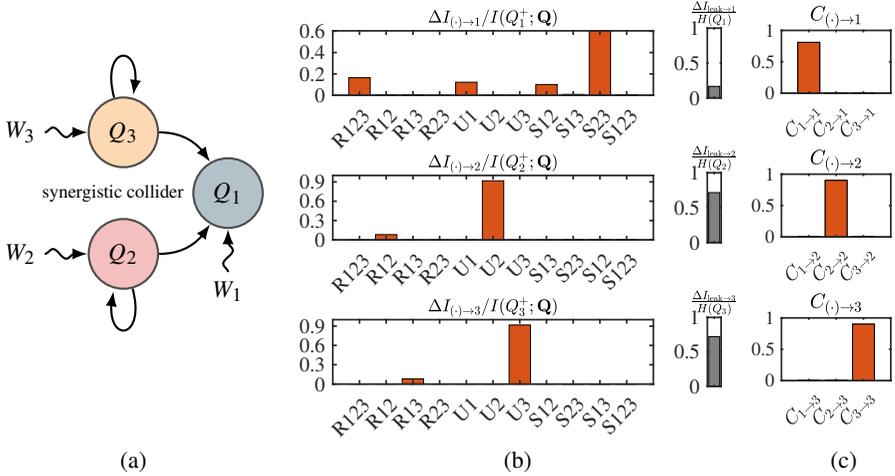
\begin{figure}
 \begin{center}
     \subfloat[]{\raisebox{1.2cm}{\begin{tikzpicture}[cir/.style={circle,draw=black!70!white,,thick,inner sep=.5em},
    >={Latex[length=.2cm]},scale=.8]
    
    \colorlet{cA}{myc2!30!white}
    \colorlet{cB}{myc1!30!white}
    \colorlet{cC}{myc3!30!white}

    \node [cir,fill=cB] (q1) at (.73,0)   {$Q_1$}; 
    \node [cir,fill=cA] (q2) at (-1,-1) {$Q_2$}; 
    \node [cir,fill=cC] (q3) at (-1,+1) {$Q_3$}; 
    
    \node [anchor=east] at (q1.west) {\scriptsize synergistic collider}; 

    \draw[decorate,decoration={snake,post length=4pt,amplitude=.2em}, thick,->]
    (-2.3,1) node[anchor=east] {$W_3$} -- (q3);
    \draw[decorate,decoration={snake,post length=4pt,amplitude=.2em}, thick,->]
    (-2.3,-1) node[anchor=east] {$W_2$} -- (q2);
    \draw[decorate,decoration={snake,post length=4pt,amplitude=.2em}, thick,->]
    (.73,-1.3) node[anchor=north] {$W_1$} -- (q1);

    \path[thick,->]
    (q3) edge[bend left] node [left] {} (q1)
    (q2) edge[bend right] node  [left] {} (q1);

    \path[thick,<-] (q2) edge[loop below] node {} (q2);
    \path[thick,<-] (q3) edge[loop above] node {} (q3);

\end{tikzpicture}}}
    \hspace{0.3cm}
      \subfloat[]{\includegraphics[width=0.43\textwidth]{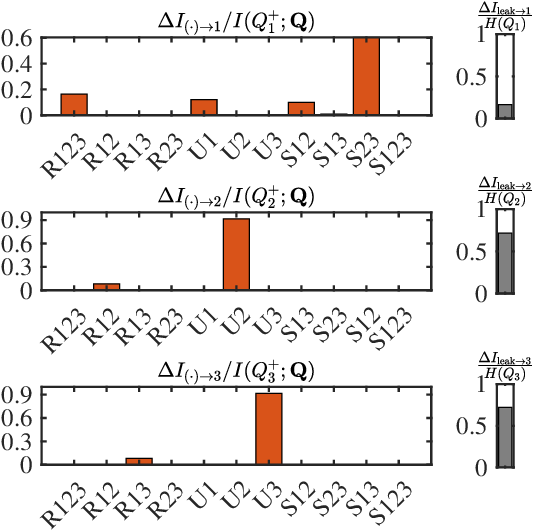}}
      \hspace{0.1cm}
      \subfloat[]{\raisebox{0.15cm}{\includegraphics[width=0.138\textwidth]{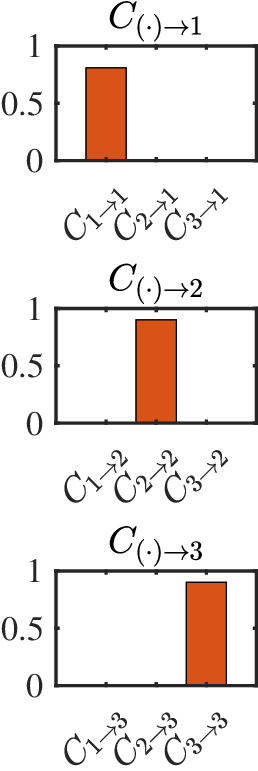}}}
\end{center}
\caption{ Collider with synergistic variables from
  Eq.~(\ref{eq:appen3}). (a) Schematic of the functional dependence
  among variables. $W_i$ represents stochastic forcing to the
  variable.  (b) Redundant, unique and synergistic causalities. The
  grey bar is the information leak.  The IT-causalities are ordered from
  left to right according to $N_{\bi \rightarrow j}^{\alpha}$.
  (c) Time cross-correlation between variables. \label{fig:appen3}}
  \end{figure}
  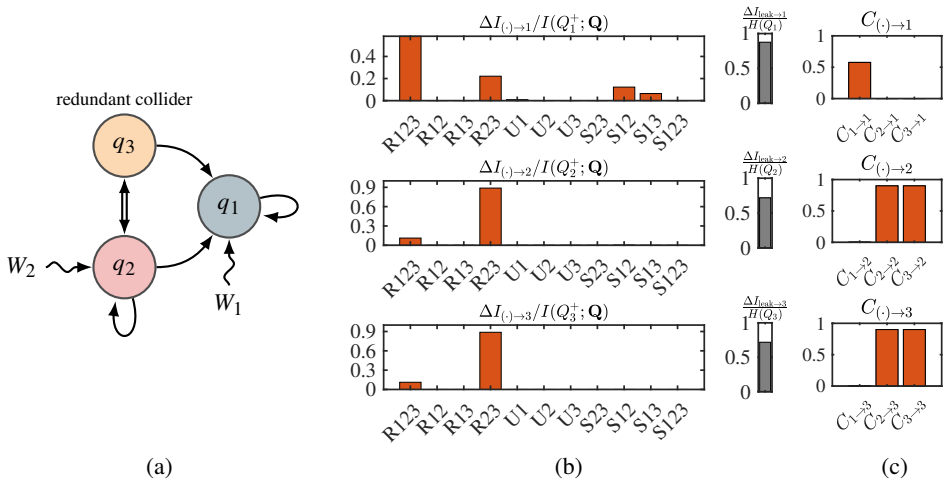
\begin{figure}
    \begin{center}
     \subfloat[]{\raisebox{1.2cm}{\begin{tikzpicture}[cir/.style={circle,draw=black!70!white,,thick,inner sep=.5em},
    >={Latex[length=.2cm]},scale=.8]
    
    \colorlet{cA}{myc2!30!white}
    \colorlet{cB}{myc1!30!white}
    \colorlet{cC}{myc3!30!white}

    \node [cir,fill=cB] (q1) at (1.73,0)   {$q_1$}; 
    \node [cir,fill=cA] (q2) at (0,-1)     {$q_2$}; 
    \node [cir,fill=cC] (q3) at (0,+1)     {$q_3$}; 

    \node[anchor=south] at (q3.north) {\scriptsize redundant collider};
    \draw[decorate,decoration={snake,post length=4pt,amplitude=.2em}, thick,->]
    (-1.3,-1) node[anchor=east] {$W_2$} -- (q2);
    \draw[decorate,decoration={snake,post length=4pt,amplitude=.2em}, thick,->]
    (1.73,-1.3) node[anchor=north] {$W_1$} -- (q1);

    \path[thick,->]
    (q3) edge[bend left] node [left] {} (q1)
    (q2) edge[bend right] node  [left] {} (q1);

    \path[thick,<-] (q2) edge[loop below] node {} (q2);
    \path[thick,<-] (q1) edge[loop right] node {} (q1);
    
    \draw[thick,double,<->,double distance=.1em] (q3) -- (q2) node [pos=.6,anchor=east] {\scriptsize };
\end{tikzpicture}}}
    \hspace{0.3cm}
      \subfloat[]{\includegraphics[width=0.43\textwidth]{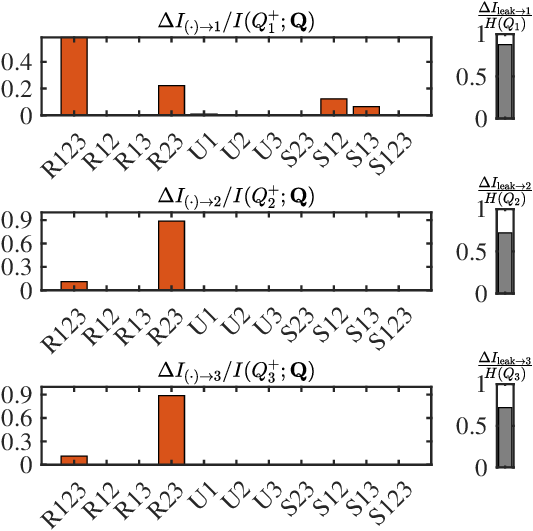}}
      \hspace{0.1cm}
      \subfloat[]{\raisebox{0.15cm}{\includegraphics[width=0.138\textwidth]{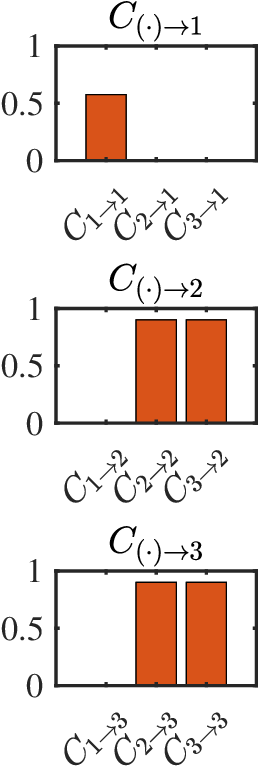}}}
    \end{center}
    \caption{ Collider with redundant variables from
      Eq.~(\ref{eq:appen5}). (a) Schematic of the functional
      dependence among variables. $W_{i}$ represents stochastic
      forcing to the variable.  (b) Redundant, unique and synergistic
      causalities. The grey bar is the information leak. The
      IT-causalities are ordered from left to right according to
      $N_{\bi \rightarrow j}^{\alpha}$.  (c) Time
      cross-correlation between variables.  \label{fig:appen5}}
  \end{figure}
 %
\item Collider with redundant variables ($Q_2 \equiv Q_3 \rightarrow
  Q_1$).  The last example corresponds to the system:
  \begin{subequations}
    \label{eq:appen5}
    \begin{align}
      Q_1(n+1) &=  0.3Q_1(n) + 0.7\{ \sin[Q_2(n)Q_3(n)] + 0.1 W_1(n)\}\\
      Q_2(n+1) &= 0.9 Q_2(n) + 0.1W_2(n), \\
      Q_3(n+1) &\equiv Q_2(n+1),
    \end{align}
  \end{subequations}
  where $Q_3$ is identical to $Q_2$. In this scenario, both $Q_2$ and
  $Q_3$ convey the same information regarding each other influence on
  the future of $Q_1$. Consistently, the non-zero IT-causalities for
  $Q_2$ and $Q_3$ are $\Delta I^R_{23 \rightarrow 2}=\Delta I^R_{23
    \rightarrow 3}\neq 0$.  The active IT-causalities in $Q_1$ is the
  redundant contribution $\Delta I^R_{123 \rightarrow 1}$ and $\Delta
  I^R_{23 \rightarrow 1}$. The variables $Q_2$ and $Q_3$ are also
  highly correlated, but no assessment can be made about their
  redundancy from the correlation viewpoint.  Furthermore, the
  correlation analysis do not show any influence from $Q_2$ and $Q_3$
  to $Q_1$.
\end{itemize}

Additional validation cases are offered in the
Appendix~\ref{sec:appendixB2}. These include coupled logistic maps
with synchronization and coupled R\"ossler-Lorenz system.

\section{Scale locality of the energy cascade in isotropic turbulence}
\label{sec:application_cascade}

The cascade of energy in turbulent flows, namely, the transfer of
kinetic energy from large to small flow scales or vice versa (backward
cascade), has been the cornerstone of most theories and models of
turbulence since the 1940s ~\citep[e.g.,][]{richardson1922,
  obukhov1941, kolmogorov1941, kolmogorov1962, aoyama2005,
  falkovich2009, cardesa2017}. However, understanding the dynamics of
kinetic energy transfer across scales remains an outstanding
challenge. Given the ubiquity of turbulence, a deeper understanding of
the energy transfer among the flow scales could enable significant
progress across various fields, ranging from combustion
~\citep{veynante2002}, meteorology ~\citep{bodenschatz2015}, and
astrophysics ~\citep{young2017} to engineering applications of
aero/hydrodynamics ~\citep{sirovich1997, hof2010, marusic2010,
  kuhnen2018, ballouz2018}. Despite the progress made in recent
decades, the causal interactions of energy among scales in the
turbulent cascade have received less attention. Here, we investigate
the redundant, unique, and synergistic causality of turbulent kinetic
energy transfer across different scales.  The primary hypothesis under
consideration here is the concept of scale locality within the
cascade, where kinetic energy is transferred sequentially from one
scale to the subsequent smaller scale.

\subsection{Numerical database}

The case chosen to study the energy cascade is forced isotropic
turbulence in a triply periodic box with side $L$.  The data were
obtained from the DNS of \citet{cardesa2015}, which is publicly
available in \citet{torroja2021}. The conservation of mass and
momentum equations of an incompressible flow are given by
\begin{eqnarray}\label{eq:cau:NS}
\frac{\partial  u_i}{\partial t} + \frac{\partial  u_i  u_j  }{\partial x_j}
 = - \frac{\partial  \Pi}{\partial x_i} +
\nu \frac{\partial^2  u_i}{\partial x_j\partial x_j} + f_i,
\quad\frac{\partial  u_i}{\partial x_i}  = 0,
\end{eqnarray}
where repeated indices imply summation, $\boldsymbol{x}=[x_1, x_2,
  x_3]$ are the spatial coordinates, $u_i$ for $i=1,2,3$ are the
velocities components, $\Pi$ is the pressure, $\nu$ is the kinematic
viscosity, and $f_i$ is a linear forcing sustaining the turbulent
flow~\citep{rosales2005}.  The flow setup is characterized by Reynolds
number based on the Taylor microscale~\citep{pope2000},
$Re_\lambda\approx 380$. The simulation was conducted by direct
numerical simulation of Eq. (\ref{eq:cau:NS}) with $1024^3$ spatial
Fourier modes, which is enough to accurately resolve all the relevant
length-scales of the flow.

In the following, we provide a summary of the main parameters of the
simulation. For more detailed information about the flow setup, the
readers are referred to \citet{cardesa2015}. The spatial and
time-averaged values of turbulent kinetic energy ($K=u_iu_i/2$) and
dissipation ($\varepsilon=2\nu S_{ij}S_{ij}$) are indicated as
$K_\mathrm{avg}$ and $\varepsilon_\mathrm{avg}$, respectively. Here,
$S_{ij} = (\partial u_i/\partial x_j + \partial u_j/\partial x_i)/2$
represents the rate-of-strain tensor.  The ratio between the largest
and smallest length scales in the problem can be quantified as
$L_\varepsilon/\eta = 1800$, where $L_\varepsilon =
K_\mathrm{avg}^{3/2}/ \varepsilon_\mathrm{avg}$ denotes the integral
length scale, and $\eta = (\nu^3/\varepsilon_\mathrm{avg})^{1/4}$
signifies the Kolmogorov length scale.  The generated data is also
time-resolved, with flow fields being stored at intervals of $\Delta t
= 0.0076 T_\varepsilon$, where $T_\varepsilon =
K_\mathrm{avg}/\varepsilon_\mathrm{avg}$. The simulation was
intentionally run for an extended period to ensure the accurate
computation of specific mutual information. The total simulated time
after transient effects was equal to $165 T_\varepsilon$.

\subsection{Characterization of the kinetic energy transfer}

\begin{figure}
  \begin{center}
    \subfloat[]{\includegraphics[width=0.45\textwidth]{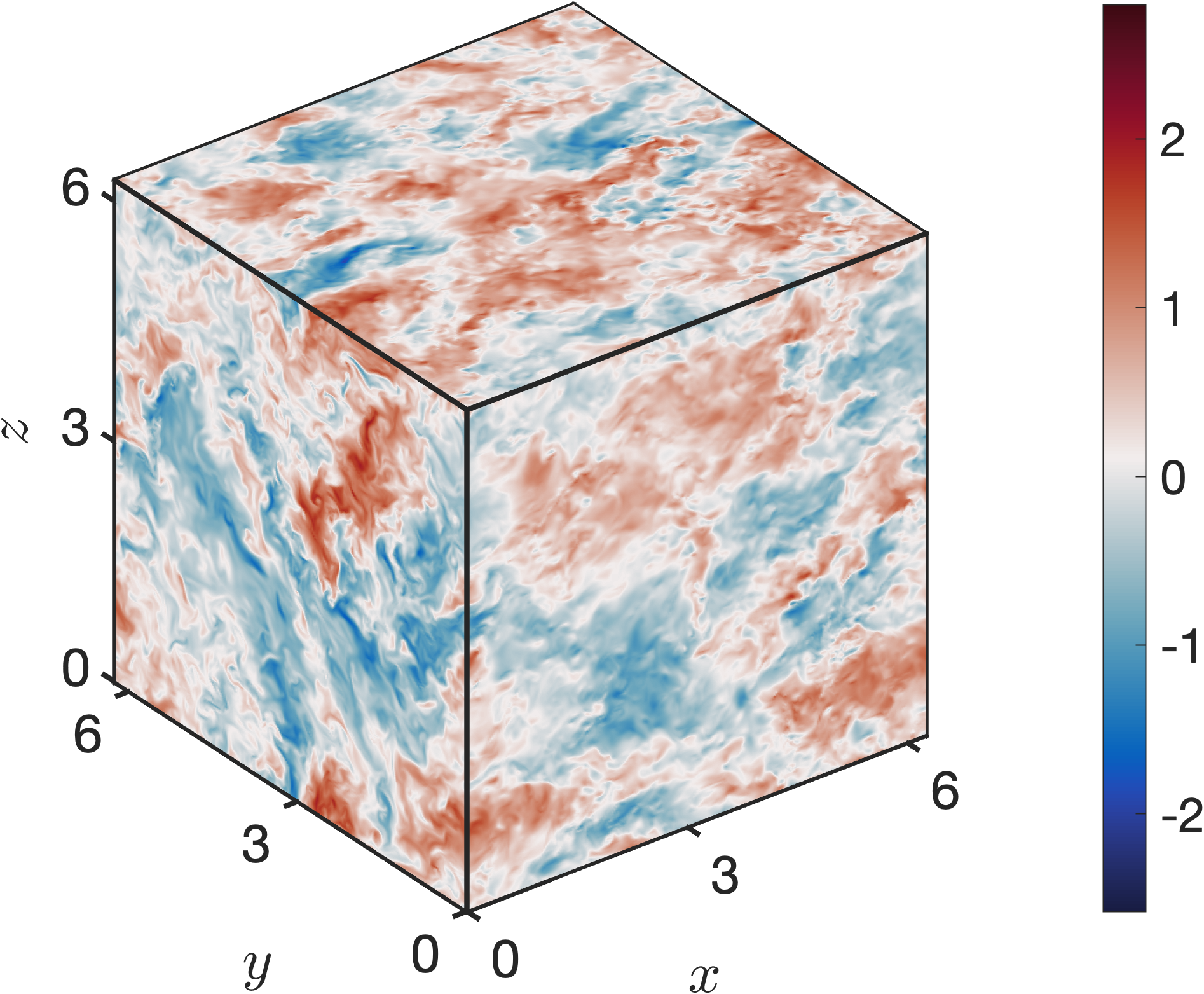}}
    \subfloat[]{\includegraphics[width=0.45\textwidth]{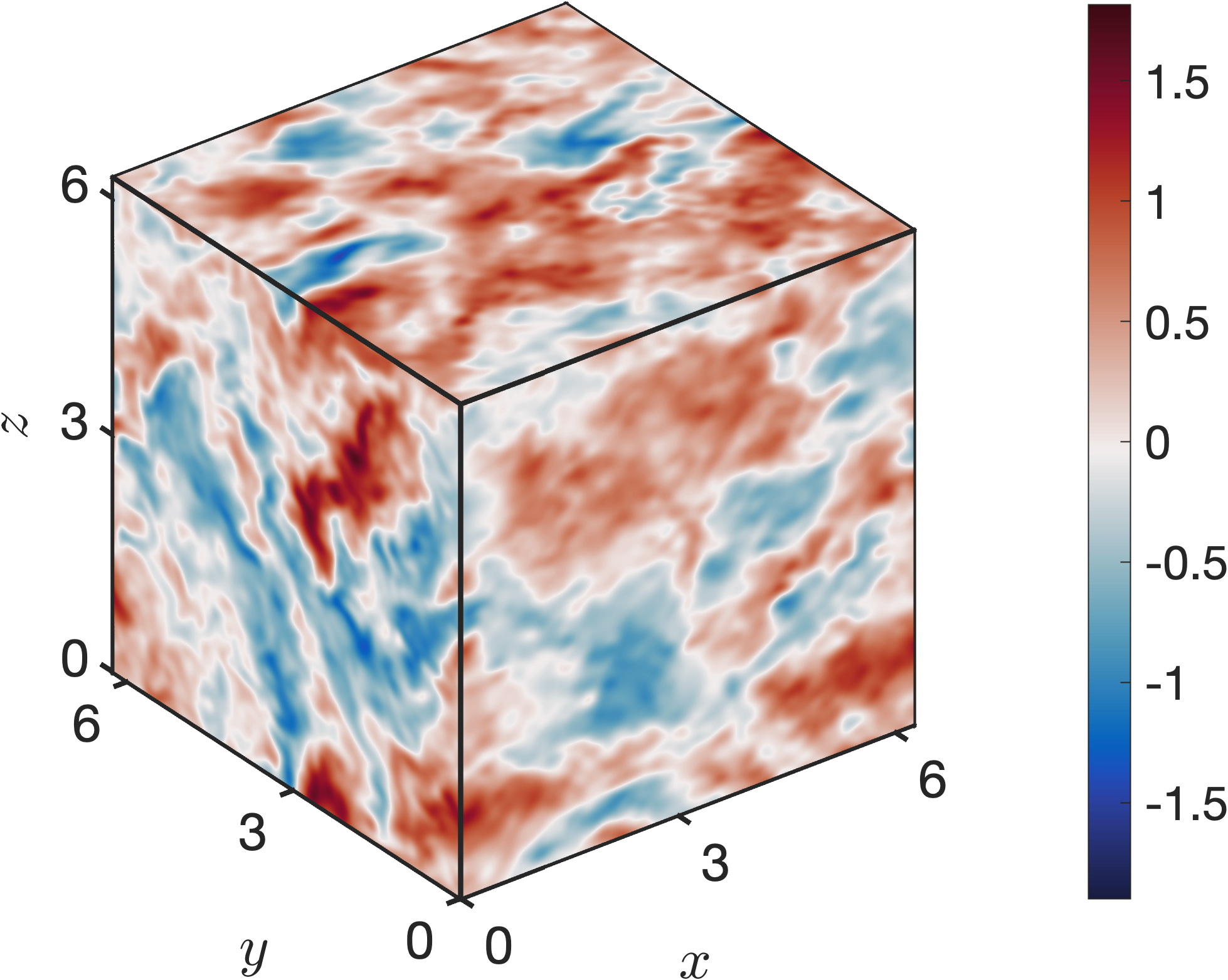}}
 \end{center}
\caption{Instantaneous velocity field for (a) $u_1$ and (b)
  $\bar{u}_1$. The velocities are normalized by their
  respective standard deviations.
  \label{fig:cau:TKE_iso}}
\end{figure}

The next stage involves quantifying the transfer of kinetic energy
among eddies at different length scales over time. To accomplish this,
the $i$-th component of the instantaneous flow velocity, denoted as
$u_i(\boldsymbol{x},t)$, is decomposed into contributions from large
and small scales according to $u_i(\boldsymbol{x},t) =
\bar{u}_i(\boldsymbol{x},t) + u'_i(\boldsymbol{x},t)$. The operator
$\bar{(\cdot)}$ signifies the low-pass Gaussian filter given by
\begin{equation}
    \bar{u}_i(\boldsymbol{x},t) = 
    \iiint_V \frac{\sqrt{\pi}}{\bar{\Delta}} \exp\left[-\pi^2( \boldsymbol{x} -
      \boldsymbol{x}')^2/\bar{\Delta}^2\right] u_i(\boldsymbol{x}',t) \mathrm{d}\boldsymbol{x}',
\end{equation}
where $\bar{\Delta}$ is the filter width and $V$ denotes integration
over the whole flow domain.  Examples of the unfiltered and filtered
velocities are included in figure \ref{fig:cau:TKE_iso}.
%
%
The kinetic energy of the large-scale field evolves as
\begin{equation}
    \left( \frac{\partial}{\partial t} + \bar{u}_j \frac{\partial }{\partial x_j}\right)\frac{1}{2}\bar{u}_i\bar{u}_i 
    =-\frac{\partial }{\partial x_j}\left( \bar{u}_j \bar{\Pi} + \bar{u}_i\tau_{ij} -  2\nu \bar{u}_i \bar{S}_{ij} \right)
    + \Sigma - 2\nu \bar{S}_{ij}\bar{S}_{ij} + \bar{u}_i \bar{f}_i, 
\end{equation}
where $\tau_{ij} = (\overline{u_i u_j} - \bar{u}_i \bar{u}_j)$ is the
subgrid-scale stress tensor, which represents the effect of the
(filtered) small-scale eddies on the (resolved) large-scale
eddies. The interscale energy transfer between the filtered and
unfiltered scales is given by
\begin{equation}
\Sigma(\bs{x},t ; \bar{\Delta}) = \tau_{ij}(\bs{x},t ; \bar{\Delta}_i) \bar{S}_{ij}(\bs{x},t ; \bar{\Delta}_i),
\end{equation}
which is the quantity of interest.

\begin{figure}
  \begin{center}
    \subfloat[]{\includegraphics[width=0.45\textwidth]{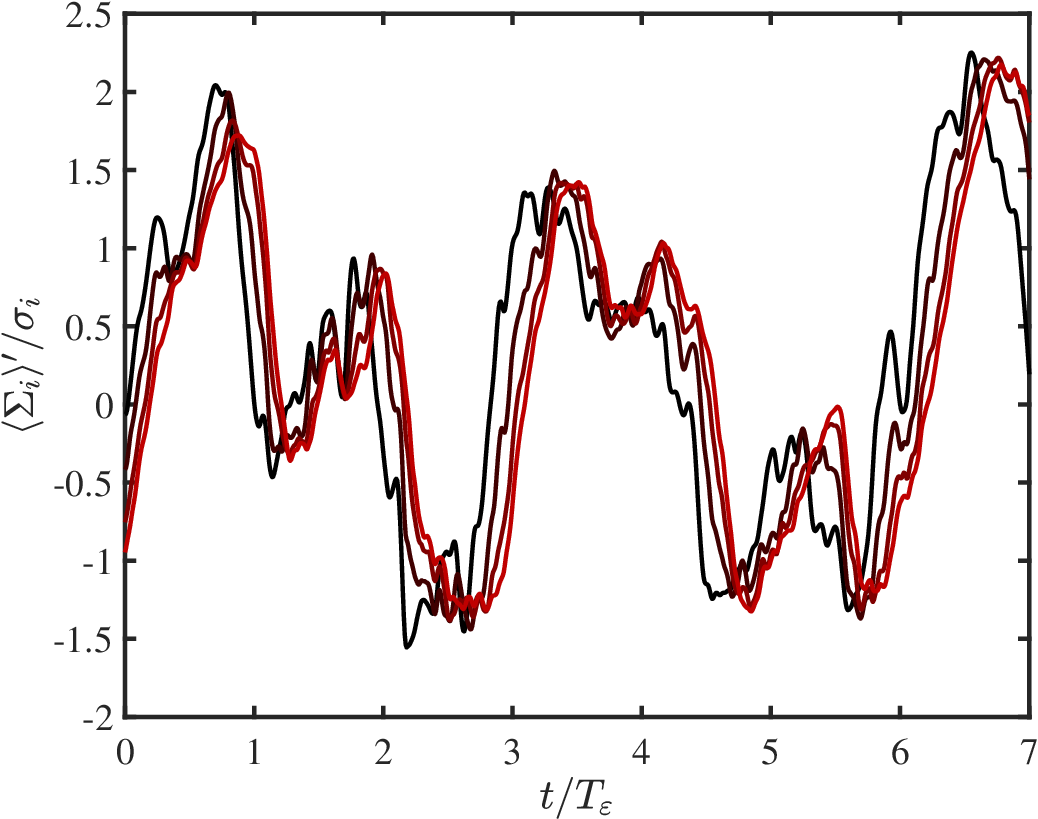}}
    \hspace{0.5cm}
    \subfloat[]{\includegraphics[width=0.45\textwidth]{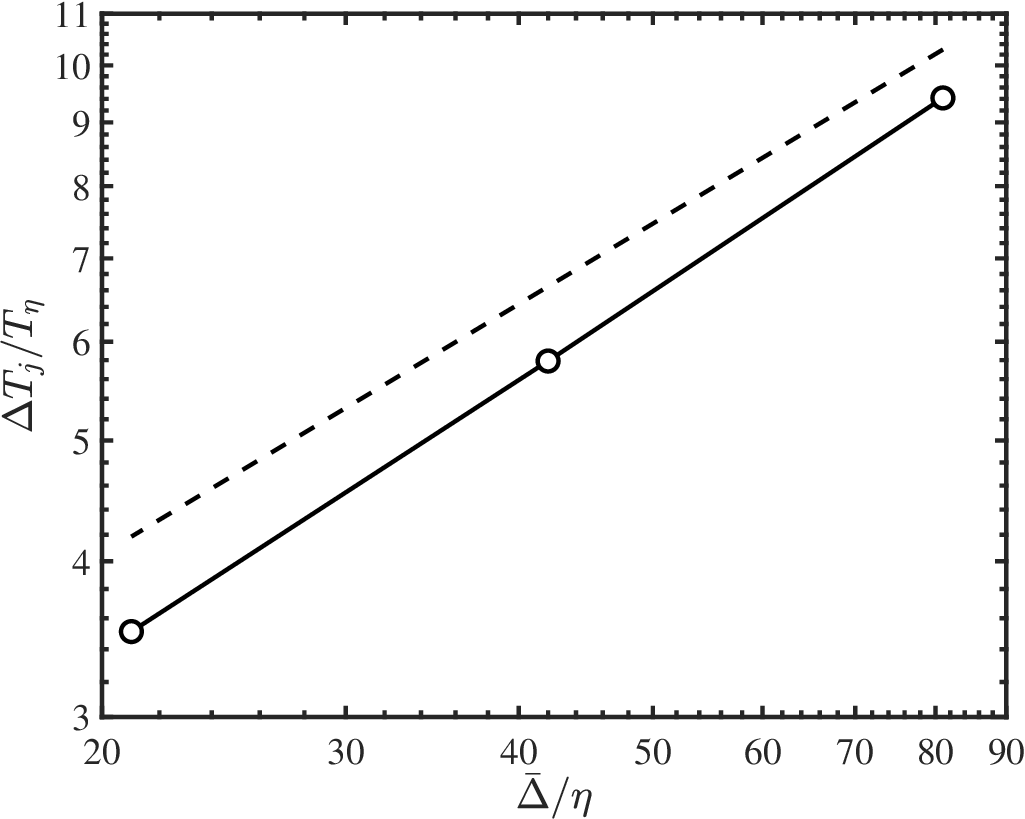}}
 \end{center}
\caption{(a) Extract of the time-history of fluctuating component of
  $\langle \Sigma_1 \rangle$, $\langle \Sigma_2 \rangle$, $\langle
  \Sigma_3 \rangle$, and $\langle \Sigma_4 \rangle$ (from black to
  red). Although not shown, the whole time-span of the signals is
  $165T_\varepsilon$.  The primes denote fluctuating component above
  the mean value and $\sigma_i$ is the standard deviation of $\langle
  \Sigma_i \rangle$. (b) Time horizon of causal influence for maximum
  unique causuality from $\langle \Sigma_i \rangle$ to $\langle
  \Sigma_j \rangle^+$. $\Delta I^U_{i \rightarrow j}$ with $j \neq i$
  as a function of the filter width. The dashed line is $\Delta T_j
  \sim \bar{\Delta}^{2/3}$. $T_\eta$ and $\eta$ are the Kolmorogov
  time-scale and length-scale, respectively.
  \label{fig:cau:signals}}
\end{figure}

The velocity field is low-pass filtered at four filter widths:
$\bar{\Delta}_1=163 \eta$, $\bar{\Delta}_2=81\eta$,
$\bar{\Delta}_3=42\eta$, and $\bar{\Delta}_4=21\eta$.  The filter
widths are situated within the inertial range of the simulation:
$L_\varepsilon > \bar{\Delta}_i > \eta$, for $i=1,2,3$ and 4. The
resulting velocity fields are used to compute the interscale energy
transfer at scale $\bar{\Delta}_i$, which is denoted by
$\Sigma_i(\bs{x},t ; \bar{\Delta}_i)$.  We use the volume-averaged
value of $\Sigma_i$ computed over the entire domain, denoted by
$\langle \Sigma_i \rangle$, as a marker for the time-evolution of the
interscale energy transfer:
\begin{equation}
    \langle \Sigma_i \rangle(t) =
    \iiint_V \Sigma(\bs{x},t ; \bar{\Delta}_i)  \mathrm{d}\boldsymbol{x},
\end{equation}
which is only a function of time for a given $\bar{\Delta}_i$. Figure
\ref{fig:cau:signals}(a) contains a fragment of the time history of
$\langle \Sigma_i \rangle$ for $i=1,2,3$ and $4$.
%

\subsection{IT-causality analysis of the energy cascade}

We examine the causal interactions among the variables representing
the interscale turbulent kinetic energy transfer: $\langle
\mathbf{\Sigma} \rangle = [\langle \Sigma_1\rangle, \langle
  \Sigma_2\rangle, \langle \Sigma_3\rangle, \langle \Sigma_4\rangle]$.
For a given target variable, $\langle \Sigma_j\rangle$, the time delay
$\Delta T_j$ used to evaluate causality is determined as the time
required for maximum $\Delta I^U_{i \rightarrow j}$ with $j \neq i$,
where $\langle \Sigma_j\rangle^+$ is evaluated at $t + \Delta
T_j$. Figure~\ref{fig:cau:signals}(b) shows that the time lags for
causal influence increase with the filter width. According to the
Kolmogorov theory~\citep{kolmogorov1941}, the characteristic lifetime
of an eddy in the inertial range scales as $\sim
\bar{\Delta}^{2/3}$. Assuming that the time required for interscale
energy transfer is proportional to the eddy lifetime, it is expected
that $\Delta T_j \sim \bar{\Delta}^{2/3}$. The values of $\Delta T_j$
are consistent with the scaling provided by $\bar{\Delta}^{2/3}$ (also
included in the figure), albeit the observation is limited to only
three scales due to low Reynolds number effects. It was tested that
the conclusions drawn below are not affected when the value of $\Delta
T_j$ was halved and doubled.

\begin{figure}
    \begin{center}
      \subfloat[]{\includegraphics[width=0.48\textwidth]{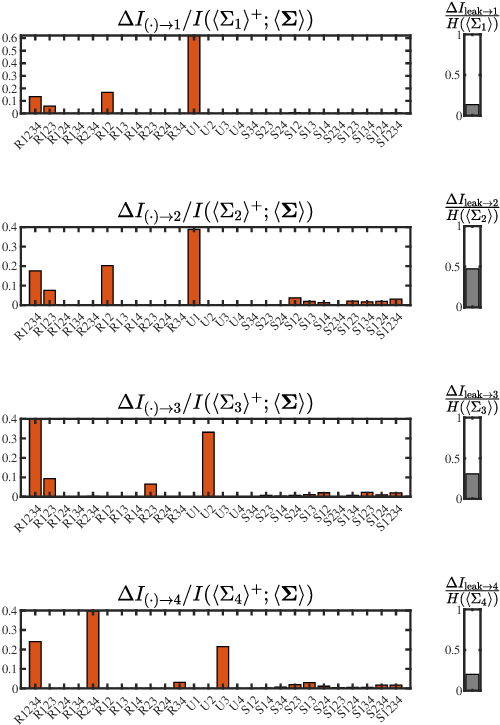}}
      \hspace{0.5cm}
      \subfloat[]{\includegraphics[width=0.4\textwidth]{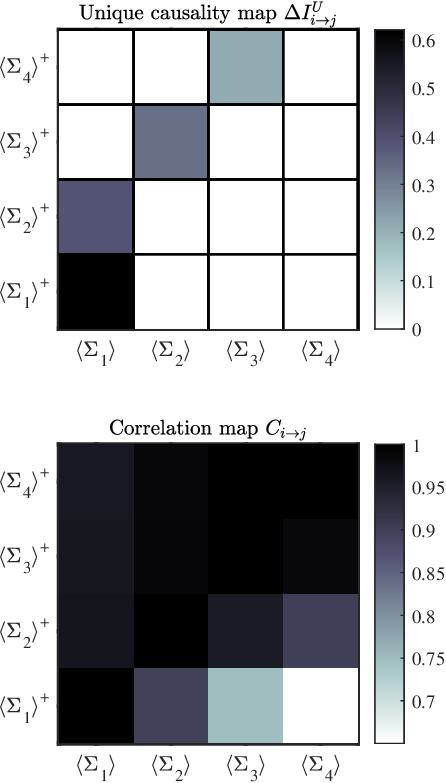}}
    \end{center}
    \caption{ (a) Redundant (R), unique (U), and synergistic (S)
      causalities among interscale energy-transfer signals at
      different scales. The information leak for each variable is also
      shown in the right-hand side bar.  The causalities are ordered
      from left to right according to $N_{\bi \rightarrow
        j}^{\alpha}$. (b) Top panel: causality maps for unique
      causalities $\Delta I^U_{i \rightarrow j}$. Bottom panel:
      Correlation map $C_{i \rightarrow j}$ between interscale
      energy-transfer signals as defined by
      Eq. (\ref{eq:Cij}). \label{fig:cau:maps1}}
\end{figure}

The redundant, unique, and synergistic causalities are shown in figure
\ref{fig:cau:maps1}(a) for each $\langle \Sigma_j\rangle$,
$j=1,\ldots,4$. The most important contributions come from redundant
and unique causalities, whereas synergistic causalities play a minor
role. The top panel in figure \ref{fig:cau:maps1}(b) shows the causal
map for $\Delta I^U_{i \rightarrow j}$. It is interesting that the
causal map for unique causalities vividly captures the forward energy
cascade of causality toward smaller scales, which is inferred from the
non-zero terms $\Delta I^U_{1\rightarrow 2}$, $\Delta
I^U_{2\rightarrow 3}$, and $\Delta I^U_{3\rightarrow 4}$. Curiously,
there is no unique causality observed from smaller to larger scales,
and any backward causality solely arises through redundant causal
relationships. In the context of IT-causality, this implies that no
new information is conveyed from the smaller scales to the larger
ones.
%

It is also revealing to compare the results in the top panel of figure
\ref{fig:cau:maps1}(b) with the time cross-correlation, as the latter
is routinely employed for causal inference by the fluid mechanics
community. The time-cross-correlation ``causality'' from $\langle
\Sigma_i\rangle$ to $\langle \Sigma_j\rangle$ is defined using
Eq.~(\ref{eq:Cij}) with $Q_i = \langle \Sigma_i\rangle$ and $Q_j =
\langle \Sigma_j\rangle$. The time lag $\Delta T_{ij}$ depends on the
pair $\langle \Sigma_i\rangle$--$\langle \Sigma_j\rangle$ and is
obtained as the time for maximum correlation. The correlation map,
$C_{i\rightarrow j}$, is shown in the bottom panel of figure
\ref{fig:cau:maps1}(b). The process portrayed by $C_{i\rightarrow j}$
is far more intertwined than its IT-causality counterpart offered in
the top panel of figure \ref{fig:cau:maps1}(b). Similarly to $\Delta
I^U_{i\rightarrow j}$, the correlation map also reveals the prevailing
nature of the forward energy cascade ($C_{i\rightarrow j}$ larger for
$j>i$). However, note that $C_{i\rightarrow j}$ is always above 0.7,
implying that all the interscale energy transfers are tightly
coupled. This is due to the inability of $C_{i\rightarrow j}$ to
compensate for mediator variables (e.g., a cascading process of the
form $\langle \Sigma_1 \rangle \rightarrow \langle \Sigma_2 \rangle
\rightarrow \langle \Sigma_3\rangle$ would result in a non-zero
correlation between $\langle \Sigma_1\rangle$ and $\langle
\Sigma_3\rangle$ via the mediator variable $\langle
\Sigma_2\rangle$). As a consequence, $C_{i\rightarrow j}$ also fails
to shed light on whether the energy is cascading locally from the
large scales to the small scales (i.e., $\langle \Sigma_1 \rangle
\rightarrow \langle \Sigma_2 \rangle \rightarrow \langle \Sigma_3
\rangle\rightarrow \langle \Sigma_4\rangle$), or on the other hand,
the energy is transferred between non-contiguous scales (e.g.,
$\langle \Sigma_1 \rangle\rightarrow \langle \Sigma_3 \rangle$ without
passing through $\langle \Sigma_2 \rangle$). We have seen that
IT-causality supports the former: the energy is transferred
sequentially.

Overall, the inference of causality based on the time
cross-correlation is obscured by the often milder asymmetries in
$C_{i\rightarrow j}$ and the failure of $C_{i\rightarrow j}$ to
account for the effects of intermediate variables. In contrast, the
causal map of unique causalities $\Delta I^U_{i \rightarrow j}$
conveys a more intelligible picture of the locality of energy
transfers among different scales. Our conclusions are consistent with
evidence from diverse approaches, such as scaling
analysis~\citep{zhou1993a, zhou1993b, eyink1995, aoyama2005,
  eyink2005, mininni2006, mininni2008, aluie2009, eyink2009}, triadic
interactions in Fourier space~\citep{domaradzki1990, domaradzki2009},
time cross-correlations~\citep{cardesa2015, cardesa2017}, and transfer
entropy in the Gledzer–Ohkitana–Yamada shell
model~\citep{materassi2014}.
  
\begin{figure}
    \begin{center}
      \subfloat[]{\includegraphics[width=0.47\textwidth]{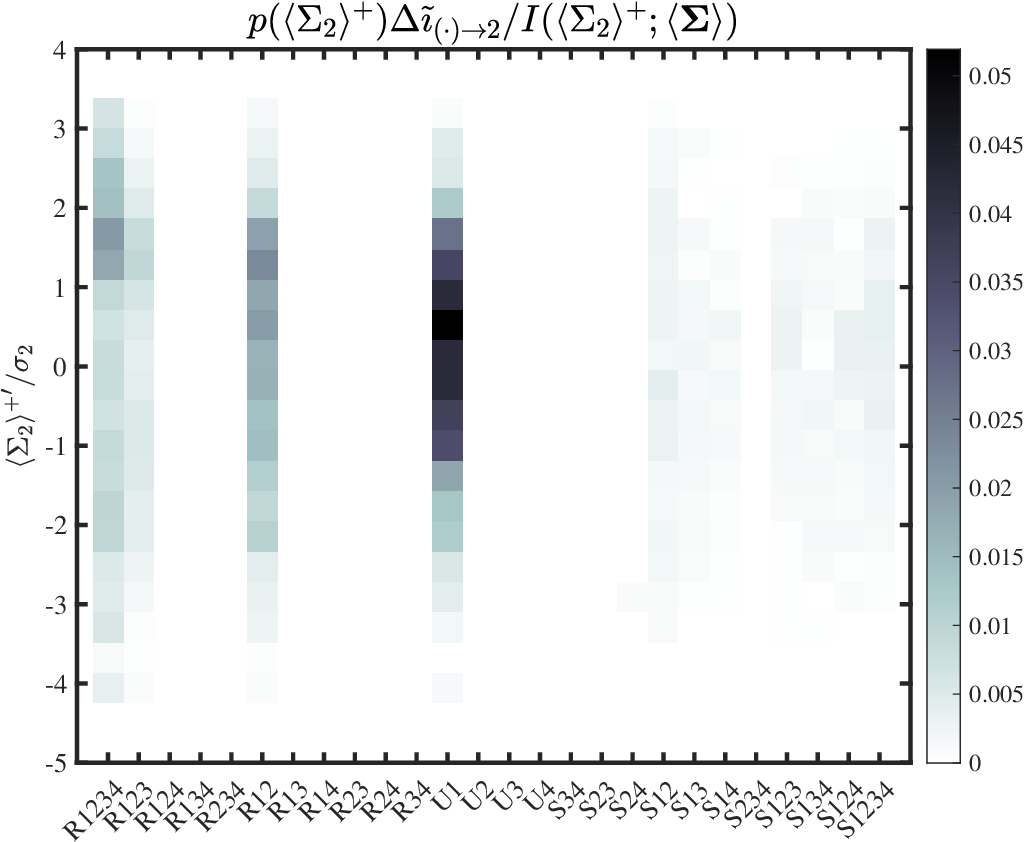}}
      \hspace{0.5cm}
      \subfloat[]{\includegraphics[width=0.47\textwidth]{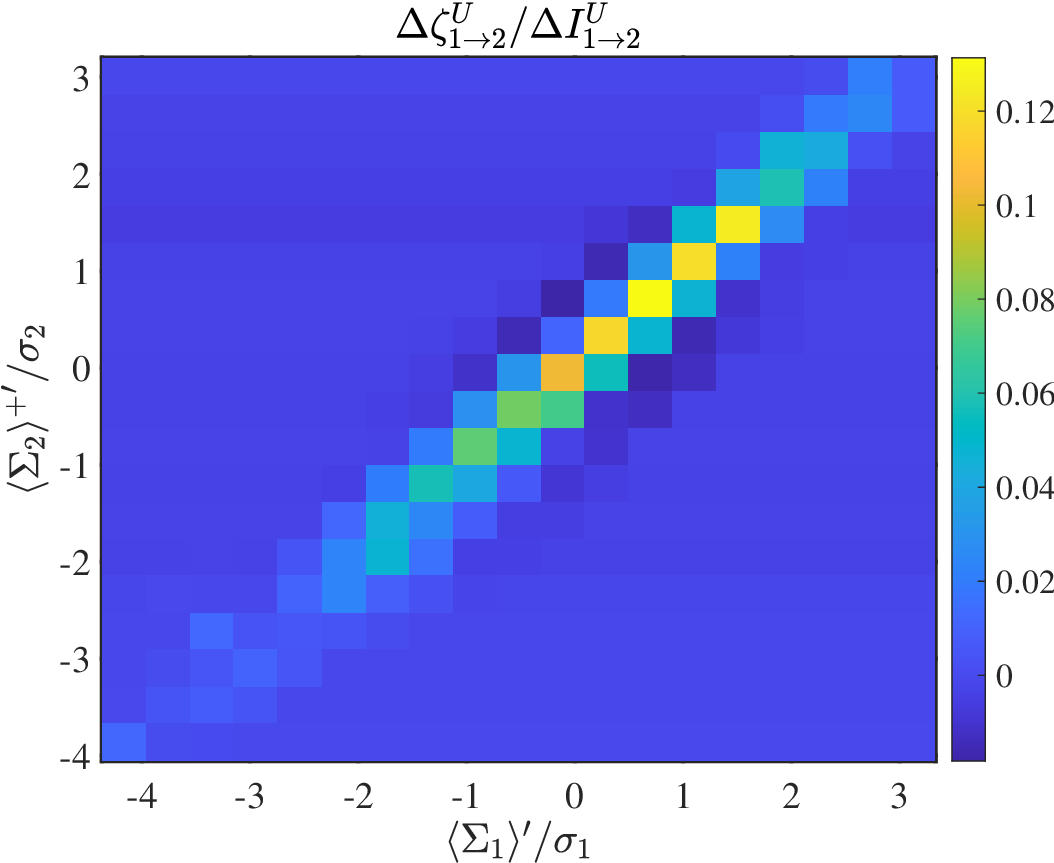}}
    \end{center}
    \caption{ Specific (a) Redundant (R), unique (U), and synergistic
      (S) causalities from $\langle \Sigma_i \rangle$ to $\langle
      \Sigma_2 \rangle^+$ as a function of the intensity of $\langle
      \Sigma_2 \rangle^+$. (b) Decomposition of unique causality as a
      function of $\langle \Sigma_1 \rangle$ and $\langle \Sigma_2
      \rangle^+$.  The primes denote fluctuating component above the
      mean value and $\sigma_i$ is the standard deviation of $\langle
      \Sigma_i \rangle$.
      \label{fig:cau:maps_intensities}}
\end{figure}

The IT-causality can be decomposed into contributions from different
intensities of the target variable ($\langle \Sigma_j
\rangle^+$). This information is provided by the weighted specific
causalities $p(\langle \Sigma_j \rangle^+) \Delta \Is_{\alpha
  \rightarrow j}^{\alpha}$, which are shown in
figure~\ref{fig:cau:maps_intensities}(a) for $\langle \Sigma_2
\rangle^+$. The unique causalities can also be decomposed as a
function of the source variable. This is done for the unique causality
from $\langle \Sigma_1 \rangle$ to $\langle \Sigma_2 \rangle^+$,
denoted by $\Delta \Iss^U_{1 \rightarrow 2}$ (as seen in
figure~\ref{fig:cau:maps_intensities}b). The results show that the
unique causality follows the linear relationship $\langle \Sigma_1
\rangle' \sim {\langle \Sigma_2 \rangle'}^+$. In general, events
located around the mean value contribute the most to the
causality. However, there is a small bias towards values above the
mean $\langle \Sigma_1 \rangle' > 0$. Similar conclusions can be drawn
for the other unique causalities (not shown).
%

Finally, we calculate the information leak ($\Delta
I_{\mathrm{leak}\rightarrow j}$) to quantify the amount of causality
unaccounted for by the observable variables. The ratios $\Delta
I_{\mathrm{leak}\rightarrow j}/H(\langle \Sigma_j \rangle)$ are found
to be 0.14, 0.47, 0.31, and 0.20 for $j=1, 2, 3$, and 4, respectively,
and for the time lags considered. The largest leak occurs for $\langle
\Sigma_2\rangle$, where approximately 47\% of the IT-causality is
carried by variables not included within the set $[\langle
  \Sigma_1\rangle, \langle \Sigma_2\rangle, \langle \Sigma_3\rangle,
  \langle \Sigma_4\rangle]$. This implies that there are other factors
affecting $\langle \Sigma_2\rangle$ that have not been accounted for
and that explain the remaining 53\% causality of the
variable. Conversely, the largest scale $\langle \Sigma_1\rangle$
bears the smallest leak of 14\%, which is due to the high value of the
unique causality $\Delta I^{U}_{1\rightarrow 1}$. The latter implies
that the future of $\langle \Sigma_1\rangle$ is mostly determined by
its own past.

\section{Interaction between inner and outer layer motions in wall-bounded turbulence}
\label{sec:application_inner-outer}

The behavior of turbulent motion within the thin fluid layers
immediately adjacent to solid boundaries poses a significant challenge
for understanding and prediction. These layers are responsible for
nearly 50\% of the aerodynamic drag on modern airliners and play a
crucial role in the first 100 meters of the atmosphere, influencing
broader meteorological phenomena~\citep{marusic2010}. The physics in
these layers involve critical processes occurring very close to the
solid boundary, making accurate measurements and simulations
exceptionally difficult.

Previous investigations utilizing experimental data from high Reynolds
number turbulent boundary layers have revealed the impact of
large-scale boundary-layer motions on the smaller-scale near-wall
cycle~\citep[e.g.][]{Hutchins2007, Mathis2009}. To date, the consensus
holds that the dynamics of the near-wall layer operate
autonomously~\citep{jimenez1999}, implying that outer-scale motions
are not necessary to sustain a realistic turbulence in the buffer
layer. Similarly, the large-scale outer-layer flow motions are known
to be insensitive to perturbations of the near-wall
cycle~\citep{townsend1976}. Nevertheless, it is widely accepted that
the influence of large-scale motions in the inner layer is manifested
through amplitude modulation of near-wall events.

In the initial studies aimed at understanding outer-inner layer
interactions~\citep{Mathis2009, marusic2010, mathis2011,
  agostini2014}, temporal or spatial signals of large-scale streamwise
velocity motions were extracted using pre-defined filter cut-offs. The
superposition of the large- and small-scale fluctuations in the
near-wall region was then parameterized assuming universal small-scale
fluctuations in the absence of inner–outer interactions. Subsequent
refinements of the method include the work by \citet{agostini2016b},
who separated the large-scale and small-scale motions by means of the
Empirical Mode Decomposition~\citep{huang1998, cheng2019} without
explicit wavelength cutoffs. Later, \citet{agostini2022} used an
auto-encoder algorithm to separate three-dimensional flow fields into
large-scale and small-scale motions. Recently, \citet{Towne2023} used
conditional transfer entropy to study inner/outer layer interactions
in a turbulent boundary layer.  \citet{Howland2018} also investigated
the small-scale response to large-scale fluctuations in turbulent
flows in a more general setting along with its implications on wall
modeling. A discussion of the amplitude modulation process in other
turbulent flows can be found in \citet{fiscaletti2015}.

\begin{figure}
\begin{center}
\includegraphics[width=1\textwidth]{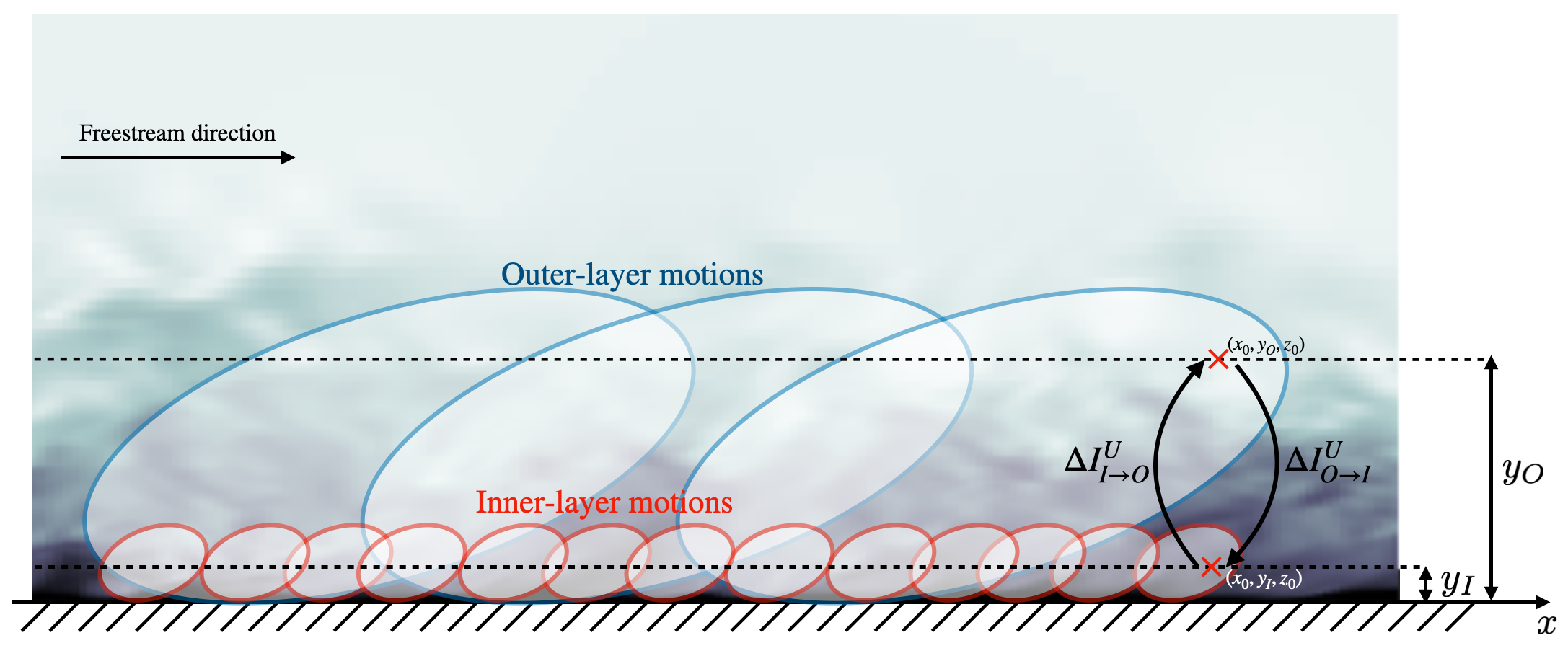}
\caption{ Schematic of outer-layer and inner-layer streamwise velocity
  motions in a turbulent boundary and their interactions via unique
  causality. \label{fig:BL:causal_sketch}}
\end{center}
\end{figure}

Here, we leverage IT-causality to investigate the interaction between
motions in the outer and inner layers of wall-bounded
turbulence. Figure \ref{fig:BL:causal_sketch} illustrates the
configuration used to examine the causal interactions between velocity
motions in the outer layer and the inner layer. Two hypotheses are
considered: (i) the footprint of outer flow large-scale motions on the
near-wall motions and (ii) Townsend's outer-layer similarity
hypothesis~\citep{townsend1976}. In hypothesis (i), the dynamics of
the near-wall motions are presumed to be influenced by outer flow
large-scale motions penetration close to the wall (i.e., predominance
of top-down causality). In hypothesis (ii), the outer region of a
turbulent boundary layer is expected to exhibit universal behavior
that is independent of the specific characteristics of the underlying
wall surface (i.e., lack of bottom-up causality).
%
%

\subsection{Numerical database}

The database utilized is a turbulent channel flow at friction Reynolds
number Re$_\tau = u_{\tau} h/\nu\approx 1,000$ from
\citet{lozano2014a} based on the channel half-height ($h$), the
kinematic viscosity ($\nu$), and the average friction velocity at the
wall ($u_{\tau}$). The simulation was generated by solving the
incompressible Navier--Stokes equations between two parallel walls via
direct numerical simulation. The size of the domain in the streamwise,
wall-normal, and spanwise direction is $2\pi h \times 2h \times \pi
h$, respectively.  Velocity fields were stored sequentially in time
every 0.8 plus units to resolve a large amount of the frequency
content of the turbulent flow. Here, plus units are defined in terms
of the friction velocity and the kinematic viscosity.  The streamwise,
wall-normal, and spanwise directions of the flow are denoted by $x$,
$y$, and $z$, respectively. More details about the simulation set-up
can be found in \citet{lozano2014a}.

\begin{figure}
\begin{center}
\includegraphics[width=0.7\textwidth]{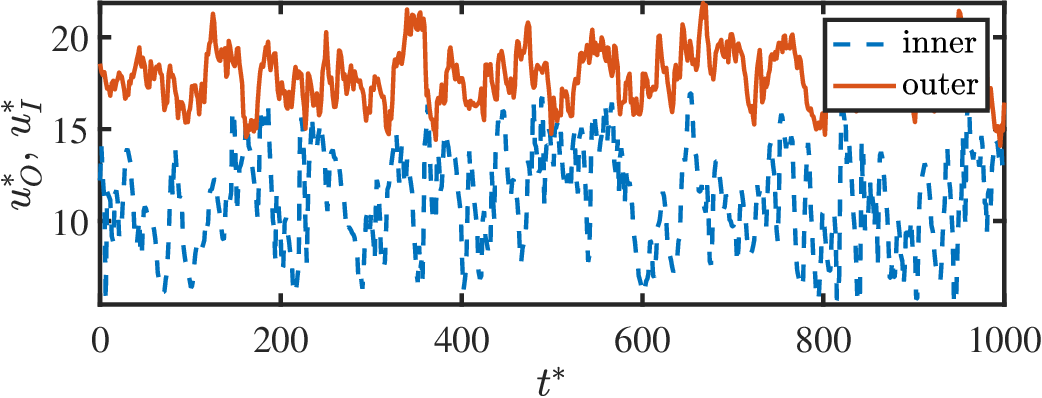}
\caption{ Example of time signals of the streamwise velocity in the
  inner layer $u_I(x,y_I,z,t)$ at $y_I^* = 15$ (dashed) and the outer
  layer $u_O(x,y_O,z,t)$ at $y_O/h=0.3$ (solid) for a fixed
  $(x,z)$. The signals are extracted from a turbulent channel flow at
  $\mathrm{Re}_\tau \approx 1,000$. The asterisk denotes plus
  units. \label{fig:BL:signals}}
\end{center}
\end{figure}

The time-resolved signals considered are those of the streamwise
velocity at the wall-normal heights $y_I^*=15$ (for the inner layer)
and $y_O/h=0.3$ (for the outer layer), where superscript $*$ denotes
plus units.  The inner and outer layer streamwise velocity signals are
denoted by $u_I(t) = u(x,y_I,z,t)$ and $u_O(t) = u(x,y_O,z,t)$,
respectively. Figure \ref{fig:BL:signals} shows an excerpt of $u_I$
and $u_O$ for a fixed $(x,z)$ location. The joint probability
distribution to evaluate IT-causality was calculated using multiple
streamwise and spanwise locations $(x, z)$. Some studies have
considered a space-shift in the streamwise direction between both
signals to increase their correlation~\citep{Howland2018}. Here, we do
not apply a streamwise space-shift between the two signals but instead
we do account for the relative displacement between signals using the
time lag $\Delta T$ to evaluate the IT-causality.
%

\subsection{IT-causality analysis of inner/outer layer interactions}

The goal is to evaluate the cause-and-effect interactions between
$u_I$ and $u_O$. Four causal interactions are investigated: the
self-induced unique causality of the signals, $\Delta
I^U_{O\rightarrow O}$ (outer$\rightarrow$outer) and $\Delta
I^U_{I\rightarrow I}$ (inner$\rightarrow$inner), and the cross-induced
unique causality between signals, $\Delta I^U_{I\rightarrow O}$
(inner$\rightarrow$outer) and $\Delta I^U_{O\rightarrow I}$
(outer$\rightarrow$inner). The latter represents the interaction
between outer and inner layer motions.

\begin{figure}
\begin{center}
  \subfloat[]{\includegraphics[width=0.47\textwidth]{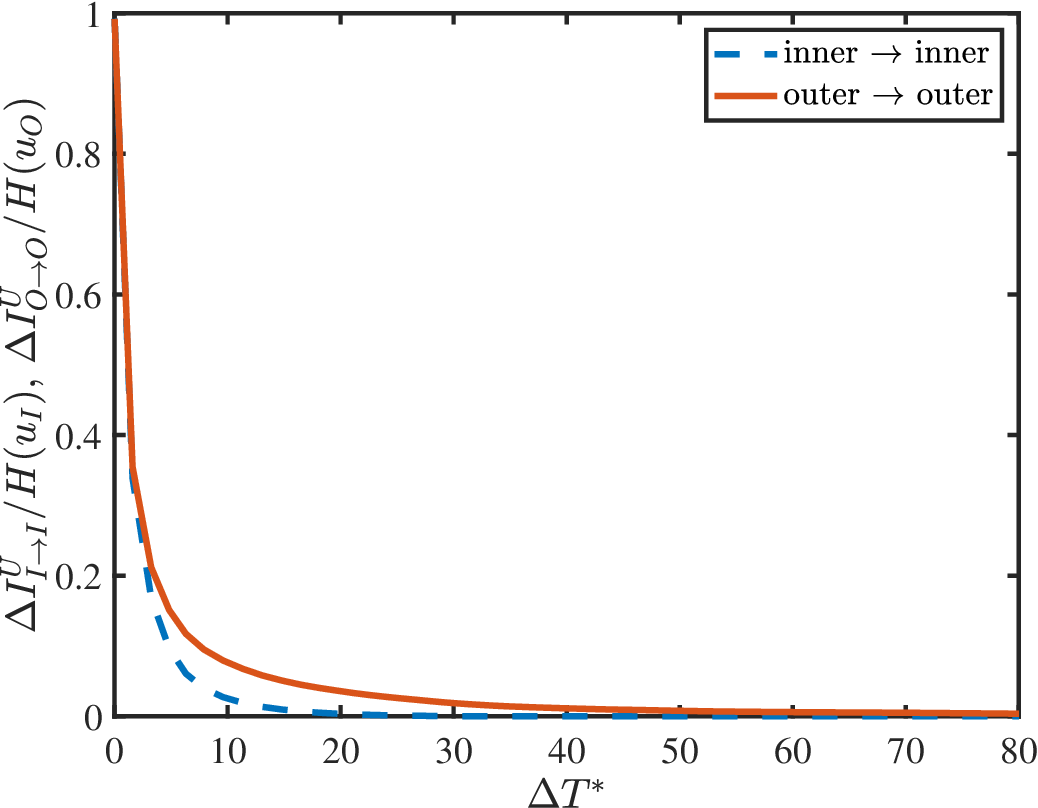}}
  \hspace{0.5cm}
  \subfloat[]{\includegraphics[width=0.47\textwidth]{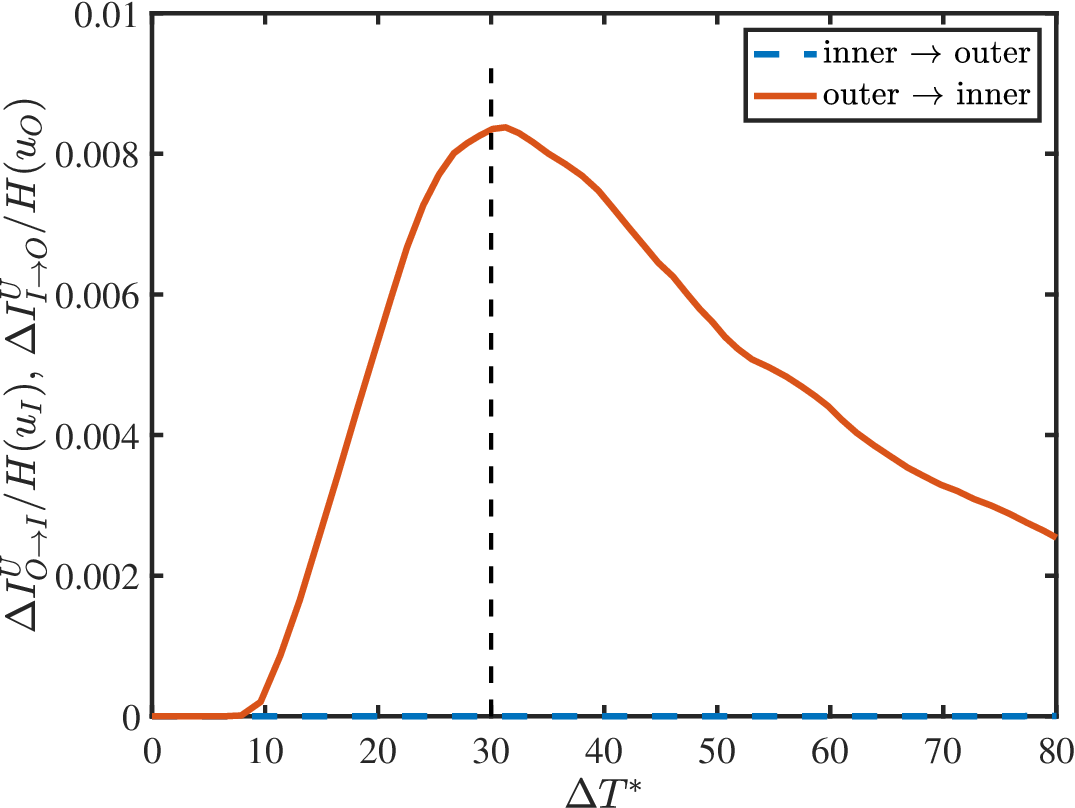}}
\end{center}
\caption{ Unique causality of the inner and outer layer streamwise velocity
motions. (a) Self- and (b) cross-induced unique causality as a
function of the time lag. The vertical dashed line is $\Delta T^* =
30$.\label{fig:BL_DT}}
\end{figure}

The self- and cross- unique causalities are compiled in figure
\ref{fig:BL_DT} as a function of the time lag $\Delta T$. The
self-induced unique causalities $\Delta I^U_{O\rightarrow O}$ and
$\Delta I^U_{I\rightarrow I}$ dominate for short time-scales. This is
expected, as variables are mostly causal to themselves in the
immediate future.  Regarding inner/outer interactions, there is a
clear peak at $\Delta T^* \approx 30$ from the outer motions to the
inner motions as seen in $\Delta I^U_{O\rightarrow I}$. This peak is
absent in the reverse direction $\Delta I^U_{I\rightarrow O}$, which
is essentially zero at all time scales. The result distinctly supports
the prevalence of top-down interactions: causality flows predominantly
from the outer-layer large-scale motions to inner-layer small-scale
motions. The outcome is consistent with the modulation of the
near-wall scales by large-scale motions reported in previous
investigations~\citep[e.g.][]{Hutchins2007, Mathis2009}. The lack of
bottom-up causality from the inner layer to the outer layer is also
consistent with the Townsend's outer-layer similarity
hypothesis~\citep{townsend1976} and previous observations in the
literature~\citep[e.g.][]{flack2005, flores2006, busse2012,
  Mizuno2013, Chung2014, lozano2019x}.
%

\begin{figure}
\begin{center}
  \subfloat[]{\includegraphics[width=0.47\textwidth]{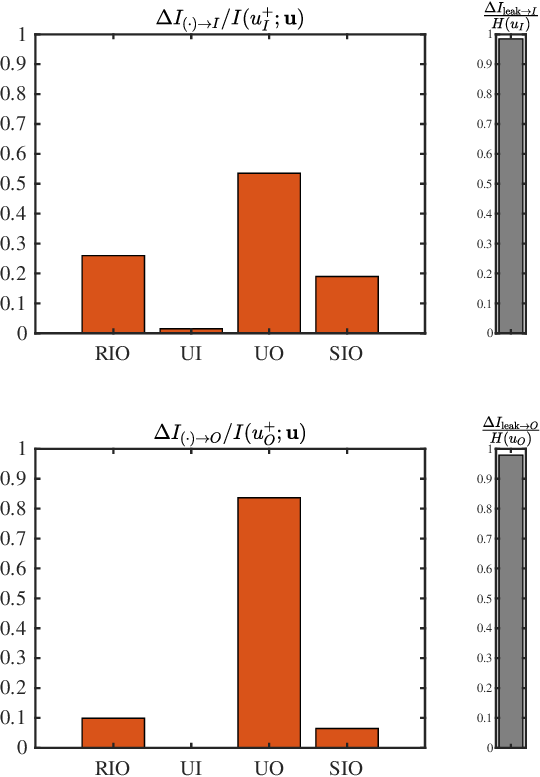}}
  \hspace{0.8cm}
  \subfloat[]{\includegraphics[width=0.37\textwidth]{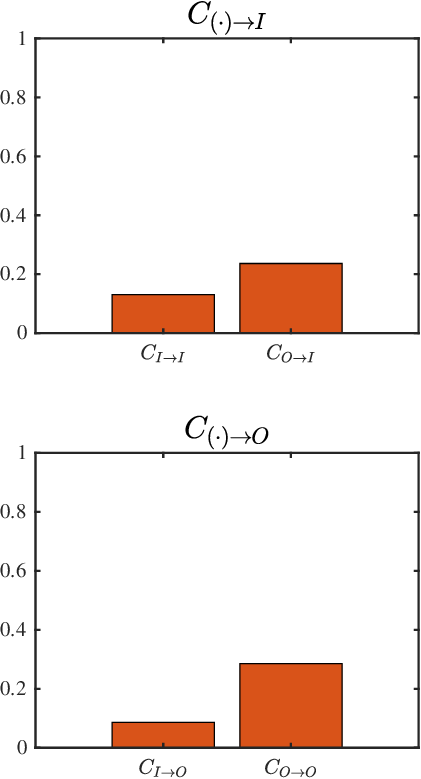}}
\end{center}
\caption{ (a) Redundant (R), unique (U), and synergistic (S)
  causalities among velocity signals in the inner (I) and outer (O)
  layer of wall-bounded turbulence. The information leak for each
  variable is also shown in the right-hand side bar. (b) Time
  cross-correlation between variables. The results are for a time-lag
  $\Delta T^* \approx 30$. \label{fig:BL_maps}}
\end{figure}

Figure \ref{fig:BL_maps}(a) shows the redundant, unique, and
synergistic causalities among velocity signals in the inner and outer
layer at $\Delta T^* = 30$. This value corresponds to the time lag of
maximum causal inference for $\Delta I^U_{O\rightarrow I}$.  The inner
layer motions are dominated by the unique causality from the outer
layer, $\Delta I^U_{O \rightarrow I}$. The redundant and synergistic
causalities are lower but still significant. Curiously, the unique
causality $\Delta I^U_{I \rightarrow I}$ is zero at the time scale
considered. For the outer-layer motions, most of the causality is
self-induced $\Delta I^U_{O \rightarrow O}$ with no apparent influence
from the inner layer.

The information leak, as indicated in the right-hand sidebar, is 99\%
for both $u_I$ and $u_O$. Such a high value implies that most of the
causality determining the future states of $u_I$ and $u_O$ is
contained in other variables not considered in the analysis. This high
information leak value is unsurprising, considering that the analysis
has neglected most of the turbulent flow field to evaluate the
causality of $u_I$ and $u_O$, retaining only the degrees of freedom
represented by two pointwise signals.

The IT-causality is contrasted with the time cross-correlations in
figure \ref{fig:BL_maps}(b) using the same time lag of $\Delta T^* =
30$. All correlations are below 30\%. Nonetheless, the fact that $C_{O
  \rightarrow I} > C_{I \rightarrow O}$ also hints at the higher
influence of the outer layer flow on the inner layer motions. However,
this asymmetry in the directionality of the interactions is milder and
less assertive than the IT-causality results. Furthermore,
correlations do not offer a detailed decomposition into redundant,
unique, and synergistic causality, nor do they account for the effect
of unobserved variables as quantified by the information leak.
%

\begin{figure}
\begin{center}
  \subfloat[]{\includegraphics[width=0.46\textwidth]{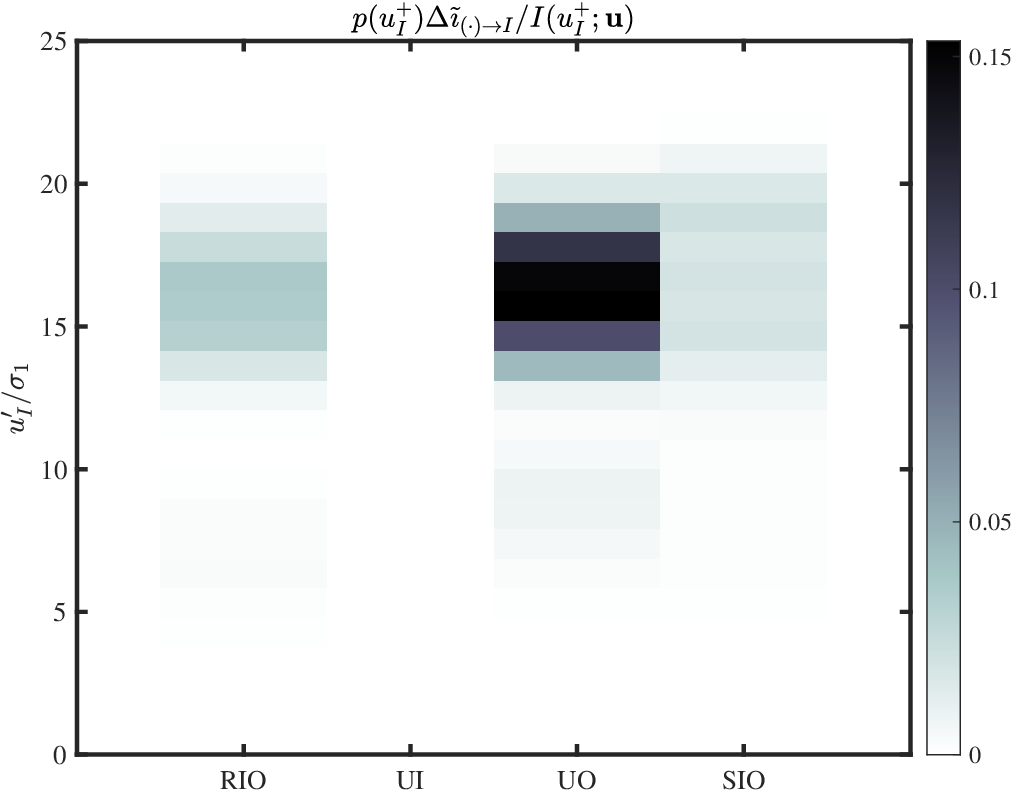}}
  \hspace{0.5cm}
  \subfloat[]{\includegraphics[width=0.48\textwidth]{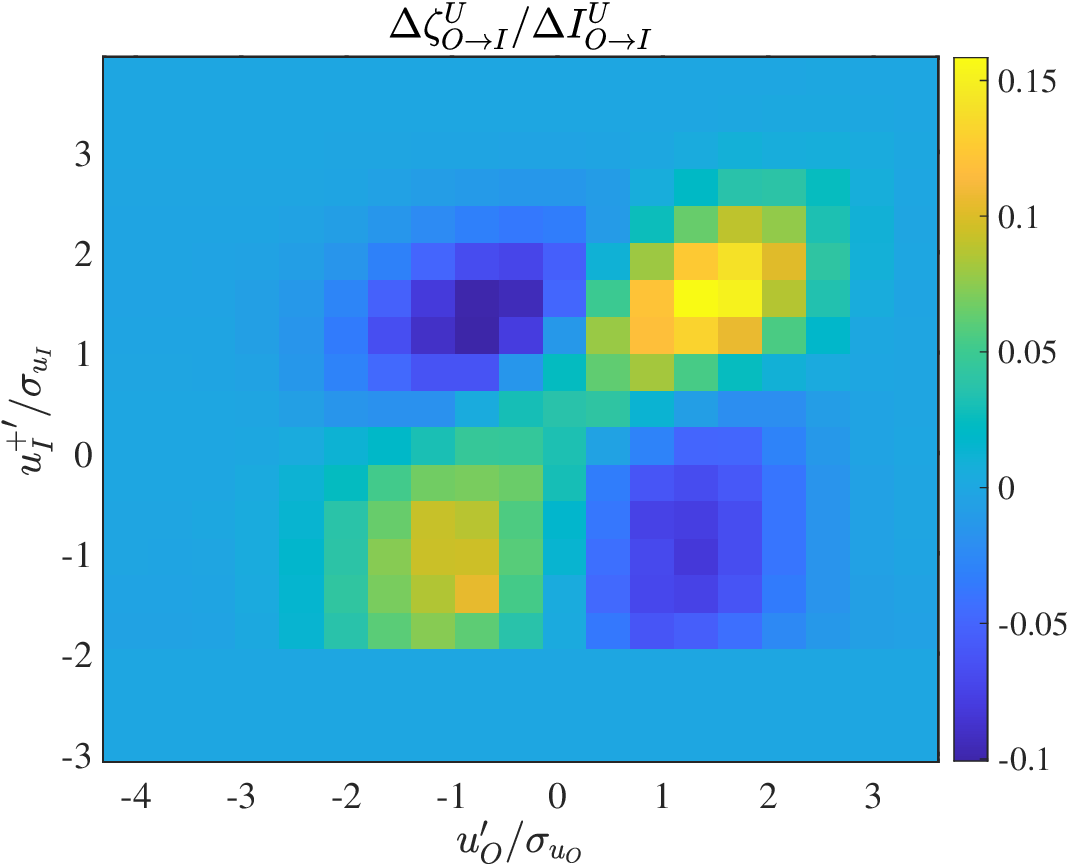}}
\end{center}
\caption{ (a) Redundant (R), unique (U), and synergistic (S) specific
  causalities from $u_O$ to $u_I^+$ as a function of the intensity of
  $u_I^+$. (b) Decomposition of unique causality $\Delta
  \Iss^U_{O \rightarrow I}$ as a function of $u_O$ to
  $u_I^+$.  The primes denote fluctuating component above the mean
  value and $\sigma_{u_I}$ and $\sigma_{u_O}$ are the standard
  deviations of $u_O$ to $u_I$, respectively.
  \label{fig:BL_intensity}}
\end{figure}

Finally, we investigate the contribution of different intensities of
$u_O$ and $u_I$ to the IT-causalities, focusing on top-down
interactions. The redundant, unique, and synergistic specific
causalities for $u_I^+$ are shown in
figure~\ref{fig:BL_intensity}(a). The unique causality as a function
of $u_O$ and $u_I$ (namely, $\Delta \Iss^U_{O \rightarrow I}$) is
presented in figure~\ref{fig:BL_intensity}(b). The contributions are
clearly divided into four quadrants. Most of the causality is located
in the first quadrant ($u_O'>0$ and $u_I'>0$), followed by the third
quadrant ($u_O'<0$ and $u_I'<0$), as both contain informative events
($\Delta \Iss^U_{O \rightarrow I}>0$). The second and fourth quadrants
contain misinformative events ($\Delta \Iss^U_{O \rightarrow I}<0$)
that do not increase the value of the unique causality. The results
seem to indicate that causality from the outer to the inner layer
primarily occurs within high-velocity streaks in the outer layer that
propagate towards the wall. A significant, albeit weaker, causality is
also observed for low-velocity streaks. It has been well-documented
that high- and low-velocity streaks are statistically accompanied by
downward and upward flow motions referred to as sweeps and ejections,
respectively~\citep{wallace2016}, which explains the high values of
$\Delta \Iss^U_{O \rightarrow I}$ for $u_O'>0$-$u_I'>0$ and
$u_O'<0$-$u_I'<0$.  The absence of causality for $u_O'$ and $u_I'$,
with opposite signs, can be attributed to the fact that these
situations are improbable, as they imply a change of sign in the
velocity streak along the wall-normal direction.
%

In summary, in comparison to previous investigations, IT-causality is
purposely devised to account for the time dynamics of the signals and
provides the time scale for maximum causal inference. Moreover, the
present approach only requires the inner/outer time signals without
any further manipulation. IT-causality is based on probability
distributions and, as such, is invariant under shifting, rescaling,
and, in general, invertible transformations of the signals (see
\S\ref{subsec:properties}). Hence, our approach unveils the
interactions between inner and outer layer velocity motions in a
simple manner while minimizing the number of arbitrary parameters
typically used in previous studies, such as the type of signal
transformation, filter width, reference convection velocity, and
selection of parameters for the extraction of the universal signal.

\section{Limitations}
\label{sec:limitations}

We conclude this work by discussing several limitations of
IT-causality for causal inference in chaotic dynamical systems:
\begin{itemize}
\item The first consideration is the acknowledgment that IT-causality
  is merely a tool. As such, it can offer valuable physical insights
  into the problem at hand when used effectively. The formulation of
  meaningful questions and the selection of appropriate variables to
  answer those questions will always take precedence over the choice
  of the tool.
\item IT-causality operates within a probabilistic framework. The
  concept of information-theoretic causality used here is based on
  transitional probabilities between states. Therefore, it should be
  perceived as a probabilistic measure of causality rather than a
  means of quantifying causality for individual events or causality
  inferred from interventions into the system.
\item IT-causality is a data-intensive tool due to the necessity of
  estimating high-dimensional probability distributions. Consequently,
  the applicability of the method is limited when dealing with a large
  number of variables. Typically, it can be applied to analyze up to
  five to ten variables simultaneously, depending on the available
  data. Attempting to consider a greater number of variables can
  introduce significant statistical errors. However, this limitation
  is expected to become less restrictive with the ongoing
  proliferation of data and advancements in probability estimation
  methods.  Appendix~\ref{sec:convergence} contains an analysis of the
  impact of the number of samples in IT-causality.
\item IT-causality relies on mutual information, which is a specific
  instance of the Kullback-Leibler divergence. Other dissimilarity
  methods exist to quantify differences between probability
  distributions, all of which are equally valid and could be explored
  in the future to devise alternative quantification of causality.
\item The results obtained through IT-causality analysis are
  contingent on the observable phase-space. High-dimensional chaotic
  systems, such as turbulence, involve nonlinear interactions among a
  large amount of variables. However, IT-causality is often limited to
  a finite subset of variables. When variables with a strong impact on
  the dynamics are unobserved or intentionally omitted due to
  practical constraints, the inferred causal structure among observed
  variables can be incomplete or misleading. This situation is
  referred to as lack of causal sufficiency. Appendix
  \S\ref{sec:Rossler} contains an example of the impact of unobserved
  variables in the coupled R{\"o}ssler-Lorenz system.
\item The concept of causality, as interpreted in IT-causality, is
  inherently linked to changes in the variables. This is reflected in
  the partitioning of the variables into states ($D_i$), which are
  regions where variable values fall within a specific range. The
  partitioning scheme shapes our definition of change; a variable is
  considered to have changed when it transitions between
  states. Hence, different partitions may lead to different
  IT-causalities. For continuous variables, IT-causality will become
  eventually insensitive to further refinements of the partition
  provided the smoothness in the probability distributions, although
  exceptions may arise. Appendix~\ref{sec:convergence} also provides
  an analysis of the sensitivity of IT-causality to partition
  refinement.
\item IT-causality is specifically designed for dynamic variables and,
  as such, cannot be employed to analyze parameters that remain
  constant over time.
\end{itemize}

In summary, IT-causality may offer valuable insights into causality
within complex systems, especially when compared to correlation-based
methods. However, like any tool, it comes with constraints that
researchers should be aware of when applying the approach to draw
meaningful conclusions.

\section{Conclusions}
\label{sec:conclusions}

Causality lies at the heart of scientific inquiry, serving as the
cornerstone for understanding how variables relate to one another. In
the case of turbulence research, the chaotic and multiscale nature of
the flow makes the quantification of causality particularly
challenging. For that reason, traditional approaches for assessing
relationships between variables often fall short in measuring causal
links, highlighting the need for more in-depth approaches.

We have introduced an information-theoretic method to quantify
causality among variables by measuring the dissimilarity between
transitional probability distributions. The approach, referred to as
IT-causality, is rooted in the forward propagation of information in
chaotic dynamical systems and quantifies causality by measuring the
information gained about future events. One distinctive feature of
IT-causality compared to correlation-based methods is its suitability
for analyzing causal networks that involve mediator, confounder, and
collider variables. In the latter, our method allows us to distinguish
between redundant, unique, and synergistic causality among
variables. IT-causality can also be decomposed as a function of
variable intensities, which facilitates the evaluation of how
different states contribute to the overall causality. Another
essential aspect of IT-causality is its foundation on probability
distributions, rendering it invariant under shifting, rescaling, and
general invertible transformations of the variables. Finally, we have
introduced the concept of information leak, quantifying the extent of
causality that remains unaccounted for due to unobserved variables.

IT-causality has been applied to investigate two problems in
turbulence research. In the first problem, we tested the hypothesis of
scale locality in the energy cascade of isotropic
turbulence. Time-resolved data from direct numerical simulations of
isotropic turbulence were used for the analysis. First, the velocity
field was low-pass filtered to obtain the interscale energy transfer
among four scales ($\bar{\Delta}$) within the inertial range. The
interscale energy transfer was volume-averaged to extract
time-resolved signals, which served as markers for the dynamics of the
energy cascade. IT-causality was applied to these signals to uncover
the causal relationships involved in the energy transfer at different
scales. It was found that the time scale for maximum causal inference
for the unique causalities follows $\bar{\Delta}^{2/3}$, consistent
with the Kolmogorov theory. Most of the causality among energy
transfer signals is either redundant or unique, with barely any
synergistic causality. This suggests that much of the information
contained in the signals is either duplicated or originates from a
unique source. In particular, the most pronounced unique causalities
occurred between consecutive scales, progressing from larger to
smaller scales. This finding supports the hypothesis of scale locality
in the interscale energy transfer in isotropic turbulence, where the
energy propagates sequentially from one scale to the next smaller
scale. Finally, the analysis of the contribution of different
intensities to the total unique causality revealed a linear
relationship between the magnitudes of the causal variable and its
effect: large-scale events of a given intensity contribute the most to
smaller-scale events of similar relative intensity.

In the second problem, we explored the interaction between streamwise
velocity motions within the inner and outer layers of wall-bounded
turbulence. To accomplish this, we utilized time-resolved data from a
direct numerical simulation of turbulent channel flow. IT-causality
was applied to two pointwise signals of the streamwise velocity. The
first signal was extracted from the outer layer, positioned at 30\% of
the channel's half-height. The second signal was located within the
inner layer, situated at a distance of 15 plus units from the
wall. The analysis revealed a unidirectional flow of causality, with
causality predominantly originating from the outer layer and
propagating towards the inner layer. This unidirectional nature
suggests a clear influence from the outer layer dynamics on those in
the inner layer, as previously observed, but not vice versa. The time
horizon for maximum causal inference from the outer to the inner layer
spanned 30 plus units. The decomposition of causality contributions as
a function of velocity intensity revealed that the causality from the
outer layer to the inner layer is primarily associated with
high-velocity streaks. These streaks extend from the outer layer down
to the wall and are consistent with the well-known association of
sweeps and high-speed streaks in wall-bounded turbulence. Lastly, it
was observed that the information leak amounted to approximately 99\%
for both velocity signals.  This substantial value indicates that a
significant portion of the causality governing the velocity signals
resides within variables that were not taken into account during the
analysis. This is expected, as most of the degrees of freedom in the
system were neglected in the analysis.

We have shown that IT-causality offers a natural approach for
examining the relationships among variables in chaotic dynamical
systems, such as turbulence. By focusing on the transitional
probability distributions of states, IT-causality provides a framework
that aligns seamlessly with the inherent unpredictability and
complexity of chaotic systems, opening up a new avenue for advancing
our understanding of these phenomena.

\section{Acknowledgements}

This work was supported by the National Science Foundation under Grant
No. 2140775 and MISTI Global Seed Funds and UPM. G.~A. was partially
supported by the Predictive Science Academic Alliance Program (PSAAP;
grant DE-NA0003993) managed by the NNSA (National Nuclear Security
Administration) Office of Advanced Simulation and Computing and the
STTR N68335-21-C-0270 with Cascade Technologies, Inc. and the Naval
Air Systems Command.  The authors acknowledge the MIT SuperCloud and
Lincoln Laboratory Supercomputing Center for providing HPC resources
that have contributed to the research results reported within this
paper. The authors would like to thank Adam A. Sliwiak, \'Alvaro
Mart\'inez-S\'anchez, Rong Ma, Sam Costa, Julian Powers, and Yuan Yuan
for their constructive comments.

\appendix

\section{Formal definition of redundant, unique, and synergistic causalities}
\label{sec:appendixA}

The problem of defining redundant, unique, and synergistic causalities
can be generally framed as the task of decomposing the mutual
information $I(Q_j^+;\bQ)$. The definitions proposed here are
motivated by their consistency with the properties presented in this
section along with the ease of interpretability.  Alternative
definitions are possible and other decompositions have been suggested
in the literature~\citep[e.g.][]{williams2010, griffith2014,
  griffith2015, ince2017, gutknecht2021, Lozano2022, kolchinsky2022}.
However, none of the previously existing decompositions are compatible
with the properties outlined in \S~\ref{subsec:properties}.

Our definitions of redundant, unique, and synergistic causalities are
motivated by the following intuition:
\begin{itemize}
\item Redundant causality from $\bQ_{\bi}=[Q_{i_1},Q_{i_2},\ldots]$ to
  $Q_j^+$ is the common causality shared among all the components of
  $\bQ_{\bi}$, where $\bQ_{\bi}$ is a subset of $\bQ$.
\item Unique causality from $Q_i$ to $Q_j^+$ is the causality from
  $Q_i$ that cannot be obtained from any other individual variable
  $Q_k$ with $k\neq i$. Redundant and unique causalities must depend
  only on probability distributions based on $Q_i$ and $Q_j^+$, i.e.,
  $p(q_i,q_j^+)$.
\item Synergistic causality from $\bQ_{\bi}=[Q_{i_1},Q_{i_2},\ldots]$
  to $Q^+_j$ is the causality arising from the joint effect of the
  variables in $\bQ_{\bi}$. Synergistic causality must depend on the
  joint probability distribution of $\bQ_{\bi}$ and $Q^+_j$, i.e.,
  $p(\bq_{\bi},q_j^+)$.
\end{itemize}

For a given state $Q_j^+=q_j^+$, the redundant, unique, and
synergistic specific information are formally defined as follows:
\begin{itemize}
\item The  specific redundant causality is the \emph{increment} in
  information gained about $q_j^+$ that is common to all the
  components of $\bQ_{\bj_k}$:
  \begin{equation}
    \Delta \Is^R_{\bj_k} =
    \begin{cases}
      \Is_{i_k}-\Is_{i_{k-1}},& \text{for} \ \Is_{i_k},\Is_{i_{k-1}} \in \GIs^1 \ \text{and} \ k\neq n_1\\
      0,              & \text{otherwise},
    \end{cases}
  \end{equation}
  where we take $\Is_{i_0}=0$, $\bj_k = [j_{k1}, j_{k2}, \ldots]$ is
  the vector of indices satisfying $\Is_{j_{kl}} \geq \Is_{i_k}$ for
  $\Is_{j_{kl}}, \Is_{i_k} \in \GIs^1$, and $n_1$ is the number of
  elements in $\GIs^1$.
  \item The specific unique causality is the \emph{increment} in
    information gained by $Q_{i_k}$ about $q_j^+$ that cannot be
    obtained by any other individual variable:
    \begin{equation}
      \Delta \Is^U_{i_k} = 
      \begin{cases}
        \Is_{i_k}-\Is_{i_{k-1}}, & \text{for}\ i_k=n_1, \ \Is_{i_{k}},\Is_{i_{k-1}} \in \GIs^1\\
        0,                 & \text{otherwise}.
      \end{cases}
    \end{equation}
  \item The specific synergistic causality is the \emph{increment} in
    information gained by the combined effect of all the variables in
    $\bQ_{\bi_k}$ that cannot be gained by other combination of
    variables $\bQ_{\bj_k}$ such that $\Is_{\bj_k} \leq \Is_{\bi_k}$
    for $\Is_{\bi_k} \in \GIs^M$ and $\Is_{\bj_k} \in
    \{\GIs^1,\ldots,\GIs^{M}\}$ with $M>1$:
    \begin{equation}
      \Delta \Is^S_{\bi_k} = 
      \begin{cases}
        \Is_{\bi_k} - \Is_{\bi_{k-1}}, & \text{for} \ \Is_{\bi_{k-1}}\geq \max\{\GIs^{M-1}\}, \ \text{and} \ \Is_{\bi_{k}}, \Is_{\bi_{k-1}} \in \GIs^M \\
        \Is_{\bi_k} - \max\{\GIs^{M-1}\}, & \text{for} \ \Is_{\bi_{k}}>\max\{\GIs^{M-1}\}>\Is_{\bi_{k-1}}, \ \text{and} \ \Is_{\bi_{k}},  \Is_{\bi_{k-1}} \in \GIs^M\\
        0,              & \text{otherwise}.
      \end{cases}
    \end{equation}
\end{itemize}  
\section{Additional validation cases}
\label{sec:appendixB2}

\subsection{Synchronization in logistic maps}

The one-dimensional logistic map is a recurrence given by
relationship,
\begin{equation}
  \label{eq:logistic_1}
  Q_1(n+1) = \alpha_1 Q_1(n)[1-Q_1(n)],  
\end{equation}
where $n$ is the time step and $\alpha_1$ is a
constant. Equation~(\ref{eq:logistic_1}) exhibits a chaotic behavior
for $\alpha_1\approx 3.57-4$~\citep{may1976}.  We consider the three
logistic maps:
\begin{subequations}
\begin{align}
  Q_1(n+1) &= \alpha_1 Q_1(n)[1-Q_1(n)],\\
  Q_2(n+1) &= \alpha_2 f_{12}[1-f_{12}],\\
  Q_3(n+1) &= \alpha_3 f_{123}[1-f_{123}],
  \end{align}
\end{subequations}
which are coupled by
\begin{subequations}
\begin{align}
  f_{12}   &= \frac{Q_2(n)+c_{1\rightarrow2}Q_1(n)}{1+c_{1\rightarrow2}},\\
  f_{123}  &= \frac{Q_3(n)+c_{12\rightarrow3}Q_1(n)+c_{12\rightarrow3}Q_2(n)}{1+2c_{12\rightarrow3}},
\end{align}
\end{subequations}
where $\alpha_1 = 3.68$, $\alpha_2 = 3.67$, and $\alpha_3 = 3.78$ are
constants, $c_{1\rightarrow2}$ is the parameter coupling $Q_2$ with
$Q_1$, and $c_{12\rightarrow3}$ is the parameter coupling $Q_3$ with
$Q_2$ and $Q_1$. The clear directionality of the variables in this
system for different values of $c_{12\rightarrow3}$ and
$c_{12\rightarrow3}$ offers a simple testbed to illustrate the
behavior of the IT-causality. The causal analysis is performed for
one time-step lag after integrating the system for $10^8$ steps. The
phase-space was partitioned using 100 bins for each variables.

\begin{figure}
  \begin{center}
    \includegraphics[width=0.32\textwidth]{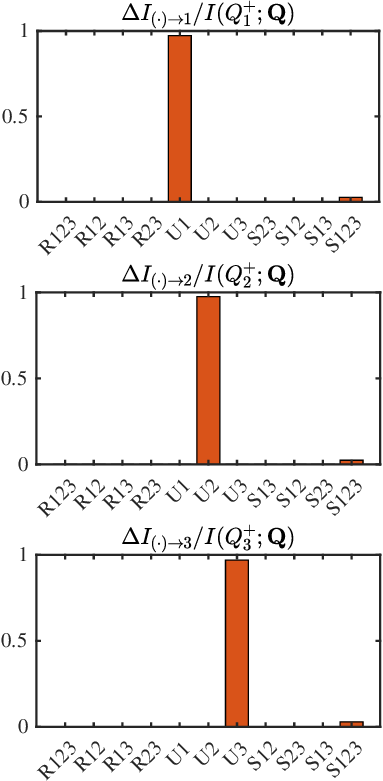}
    \includegraphics[width=0.32\textwidth]{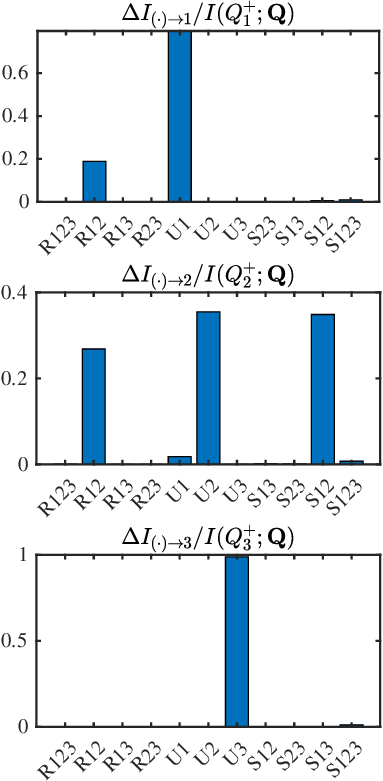}
    \includegraphics[width=0.32\textwidth]{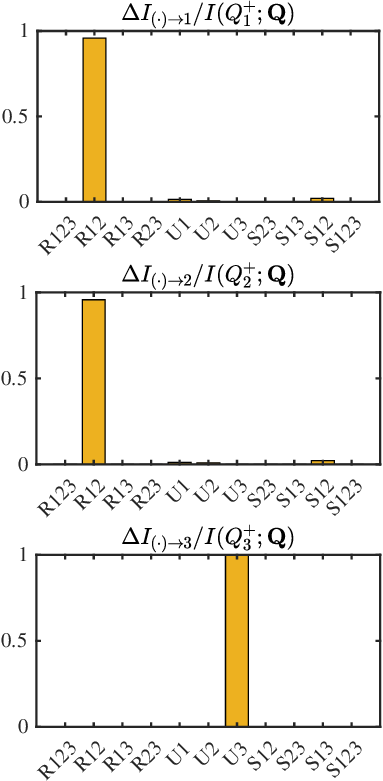}
  \end{center}
\caption{Redundant (R), unique (U) and synergistic (S) causalities for
  $Q_1$, $Q_2$, $Q_3$ in coupled logistic maps for (left panels, in
  red) uncoupled variables, $c_{12}=0$ and $c_{123}=0$, (center
  panels, in blue) intermediate coupling $Q_1 \rightarrow Q_2$,
  $c_{12}=0.1$ and $c_{123}=0$ and (right panels, in yellow) strong
  coupling $Q_1 \rightarrow Q_2$ $c_{12\rightarrow3}=1$ and
  $c_{12\rightarrow3}=0$. \label{fig:logistic_12}}
\end{figure}

First, we consider three cases with different degrees of coupling
between $Q_1$ and $Q_2$ while maintaining $Q_3$ uncoupled. The results
are shown in figure~\ref{fig:logistic_12}.
\begin{itemize}
\item Uncoupled systems
  ($c_{12\rightarrow3}=c_{12\rightarrow3}=0$). In this case, $Q_1$,
  $Q_2$, and $Q_3$ are completely uncoupled and the only non-zero
  causalities are the self-induced unique components $\Delta
  I^U_{1\rightarrow 1}$, $\Delta I^U_{2\rightarrow 2}$, and $\Delta
  I^U_{3\rightarrow 3}$, as shown by the left panels in
  figure~\ref{fig:logistic_12} (red bars).
\item Intermediate coupling $Q_1 \rightarrow Q_2$
  ($c_{12\rightarrow3}=0.1$ and $c_{12\rightarrow3}=0$). In this case,
  the dynamics of $Q_2$ are affected by $Q_1$. This is shown in the
  center panels of figure~\ref{fig:logistic_12} (blue bars) by the
  non-zero terms $\Delta I^R_{12\rightarrow 2}\neq 0$, $\Delta
  I^U_{1\rightarrow 1}\neq 0$ and $\Delta I^S_{12\rightarrow 1}\neq
  0$. The latter is the synergistic causality due to the combined
  effect of $Q_1$ and $Q_2$, which is a manifestation of the coupling
  term $f_{1\rightarrow 2}$. We can also observed that $\Delta
  I^R_{12\rightarrow 1}\neq 0$. Note that this is a redundant
  causality and does not necessarily imply that $Q_1$ is affected by
  $Q_2$. Instead, it should be interpreted as $Q_2$ being able to
  inform about the future of $Q_1$, which is expected as $Q_1$ is
  contained in the right-hand side of the equation for $Q_2$.  As
  expected, the only non-zero causality for $Q_3$ is again $\Delta
  I^U_{3\rightarrow 3}$, as it is uncoupled from $Q_1$ and $Q_2$.
\item Strong coupling $Q_1 \rightarrow Q_2$ ($c_{12\rightarrow3}=1$
  and $c_{12\rightarrow3}=0$). Taking the limit
  $c_{12\rightarrow3}\rightarrow \infty$, it can be seen that $Q_2
  \equiv Q_1$. It is also known that even for lower values of
  $c_{12\rightarrow3}\sim1$, $Q_1$ and $Q_2$ synchronize and both
  variables exhibit identical dynamics~\citep{diego2019}. This is
  revealed by the right panels of figure~\ref{fig:logistic_12} (yellow
  bars), where the only non-zero causalities are $\Delta
  I^R_{12\rightarrow 1}$ and $\Delta I^R_{12\rightarrow 2}$. As in the
  two previous cases, $Q_3$ remains unaffected ($\Delta
  I^U_{3\rightarrow 3} \neq 0$).
\end{itemize}
%

\begin{figure}
  \begin{center}
    \includegraphics[width=0.32\textwidth]{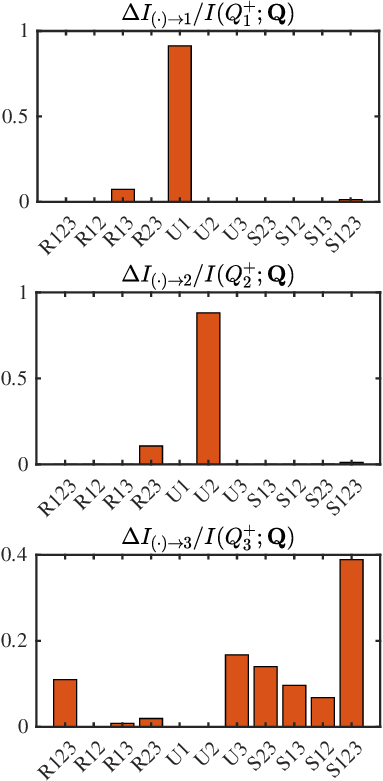}
    \includegraphics[width=0.32\textwidth]{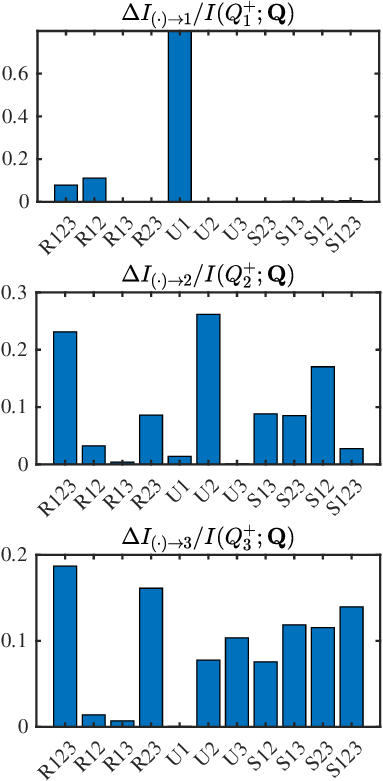}
    \includegraphics[width=0.32\textwidth]{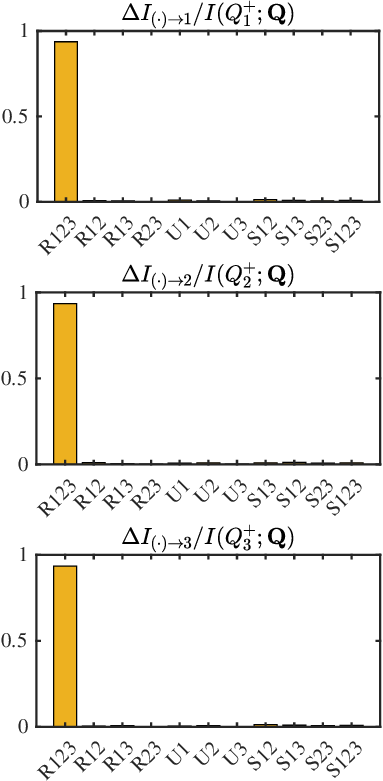}
  \end{center}
\caption{ Redundant (R), unique (U) and synergistic (S) causalities
  for $Q_1$, $Q_2$, $Q_3$ in coupled logistic maps for (left panels,
  in red) uncoupled variables, $c_{12}=0$ and $c_{123}=1$, (center
  panels, in blue) intermediate coupling $Q_1 \rightarrow Q_2$,
  $c_{12}=0.1$ and $c_{123}=0.1$ and (right panels, in yellow) strong
  coupling $Q_1 \rightarrow Q_2$, $c_{12\rightarrow3}=1$ and
  $c_{12\rightarrow3}=1$. \label{fig:logistic_123}}
\end{figure}

Next, we consider three additional cases in which $Q_3$ is coupled
with $Q_1$ and $Q_2$. The results are shown in
figure~\ref{fig:logistic_123}.
\begin{itemize}
\item Strong coupling $Q_2,Q_1\rightarrow Q_3$ and no coupling
  $Q_1\rightarrow Q_2$ ($c_{12\rightarrow3}=0$ and
  $c_{123\rightarrow3}=1$). The results, included in the left panels
  of figure~\ref{fig:logistic_123} (red bars), show that most of the
  causality to $Q_1$ and $Q_2$ is self-induced and unique ($\Delta
  I^U_{1\rightarrow 1}\neq0$ and $\Delta I^U_{2\rightarrow 2}\neq0$,
  respectively), with a small redundant contribution from $Q_3$. This
  is consistent the fact that the dynamics of $Q_1$ and $Q_2$ do not
  depend on any other variable than themselves, but $Q_3$ is coupled
  with $Q_1$ and $Q_2$, which results in small amount of redundant
  causality.  There is a strong causality from $Q_1$ and $Q_2$ to
  $Q_3$ in the form of synergistic causality, being $\Delta I^S_{123
    \rightarrow 3}$ the dominant component consistent with the
  coupling term $f_{123}$.
\item Intermediate coupling $Q_2,Q_1\rightarrow Q_3$ and
  $Q_1\rightarrow Q_2$
  ($c_{12\rightarrow3}=c_{123\rightarrow3}=0.1$). This is the most
  complex scenario since the variables do not synchronize yet they
  affect each other notably. The results are shown in the center
  panels of figure~\ref{fig:logistic_123} (blue bars). The causalities
  to $Q_1$ remain mostly independent from $Q_2$ and $Q_3$ except for
  the expected small redundant causalities. The causalities to $Q_2$
  and $Q_3$ exhibit a much richer behavior, with multiple redundant and
  synergistic causalities. The fact that $Q_2$ is not coupled to $Q_3$
  can be seen from the lack of unique causality $\Delta
  I^U_{3\rightarrow 2}$.
\item Strong coupling $Q_2,Q_1\rightarrow Q_3$ and $Q_1\rightarrow
  Q_2$ ($c_{12\rightarrow3}=1$ and $c_{123\rightarrow3}=1$).  In this
  case, the three variables synchronize such that $\Delta
  I^R_{123\rightarrow 1} = \Delta I^R_{123\rightarrow 2} = \Delta
  I^R_{123\rightarrow 3}\neq 0$ (i.e., they can be interpreted as
  exact copies of each other). The results are shown in right panels
  of figure~\ref{fig:logistic_123} (yellow bars).
\end{itemize}
%

\subsection{Example in coupled R{\"o}ssler-Lorenz system}
\label{sec:Rossler}

We study a coupled version of the Lorenz system~\citep{lorenz1963} and
the R\"ossler system~\citep{rossler1977}. The former was developed by
Lorenz as a simplified model of viscous fluid flow. R\"ossler proposed
a simpler version of the Lorenz's equations in order to facilitate the
study its chaotic properties.  The governing equations are
\begin{subequations}
\begin{align}  
  \label{eq:RL}
 \frac{\dd Q_1}{\dd t} &= -6[ Q_2 + Q_3 ], \\
 \frac{\dd Q_2}{\dd t} &= 6[Q_1 + 0.2Q_2 ], \\
 \frac{\dd Q_3}{\dd t} &= 6\left[ 0.2+Q_3[Q_1-5.7] \right], \\
 \frac{\dd Q_4}{\dd t} &= 10[ Q_5 - Q_4 ], \\
 \frac{\dd Q_5}{\dd t} &= Q_4 [28 - Q_6 ] - Q_5 + c Q_2^2, \\
 \frac{\dd Q_6}{\dd t} &= Q_4 Q_5 - \frac{8}{3}Q_6,
\end{align}
\end{subequations}
where $[Q_1,Q_2,Q_3]$ correspond to the R\"ossler system and
$[Q_4,Q_5,Q_6]$ to the Lorenz system. The coupling is unidirectional
from the R\"ossler system to the Lorenz system via $ Q_2 \rightarrow
Q_5$ and the parameter $c$. This coupled system has previously been
studied by \citet{Quiroga2000} and \citet{Krakovska2018}.

We use this case to study the behavior of IT-causality among four
variables in a continuous dynamical system when some of the variables
are hidden. The observable variables are $\bQ = [Q_1, Q_2, Q_5,
  Q_6]$. The system was integrated for $10^6 t_{\text{ref}}$ where
$t_{\text{ref}}$ is the time-lag for which $I(Q_1^+ ; Q_1)/I(Q_1 ;
Q_1) = 0.5$. The time-lag selected for causal inference is $\Delta T
\approx t_{\text{ref}}$ and the 50 bins per variable were used to
partition the phase space.

\begin{figure}
  \begin{center}
    \includegraphics[width=0.35\textwidth]{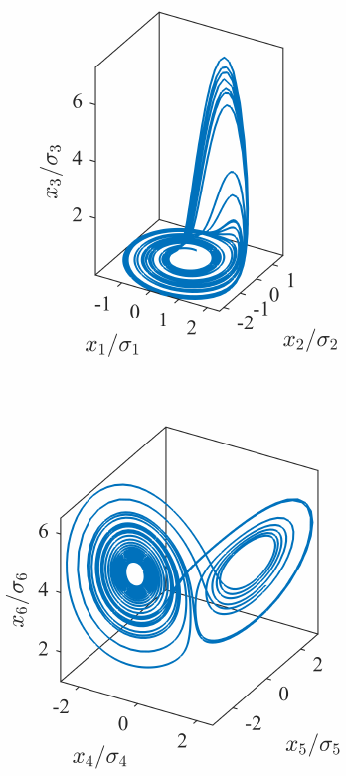}
    \hspace{0.5cm}
    \includegraphics[width=0.55\textwidth]{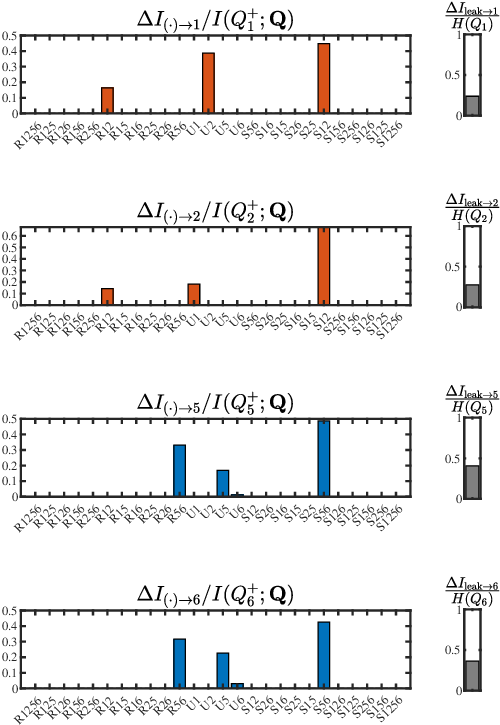}
  \end{center}
\caption{ Uncoupled R\"ossler-Lorenz system ($c=0$). The left panels
  show excerpts of the trajectories pertaining to R\"ossler systems
  $[Q_1,Q_2,Q_3]$ (top) and Lorenz system $[Q_4,Q_5,Q_6]$
  (bottom). The right panels show the redundant (R), unique (U), and
  synergistic (S) causalities among $[Q_1, Q_3, Q_4, Q_6]$. The
  causalities are ordered from left to right according to $N_{\bi
    \rightarrow j}^{\alpha}$.  \label{fig:Rossler_c0}}
\end{figure}

The results for uncoupled systems ($c=0$) are shown in figure
\ref{fig:Rossler_c0}. The left panel portrays a typical trajectories
of the systems. The causalities are shown in the right panel, where
red and blue colors are used to represent causalities exclusive to the
R\"ossler system (i.e., only involving $Q_1$ and $Q_2$) and Lorenz
system (i.e., only involving $Q_5$ and $Q_6$),
respectively. Unsurprisingly, IT-causality shows that both systems are
uncoupled. Moreover, the unique, redundant and synergistic causal
structure identified in the R\"ossler and Lorenz systems are
consistent with structure of Eq.~(\ref{eq:RL}). The information leak
is roughly 25\% due to the unobserved variables and the uncertainty
introduced by partitioning the observable phase-space.
%

\begin{figure}
  \begin{center}
    \includegraphics[width=0.35\textwidth]{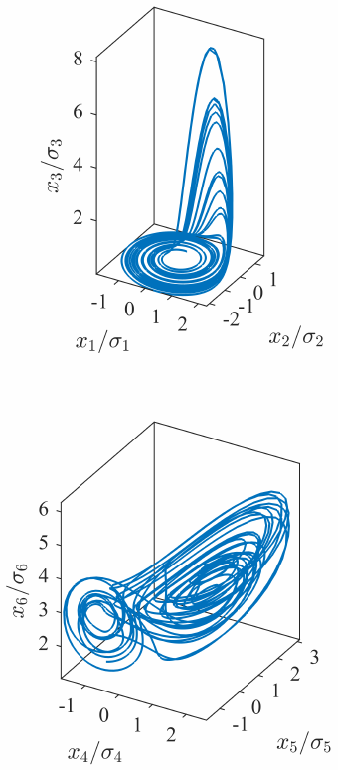}
    \hspace{0.5cm}
    \includegraphics[width=0.55\textwidth]{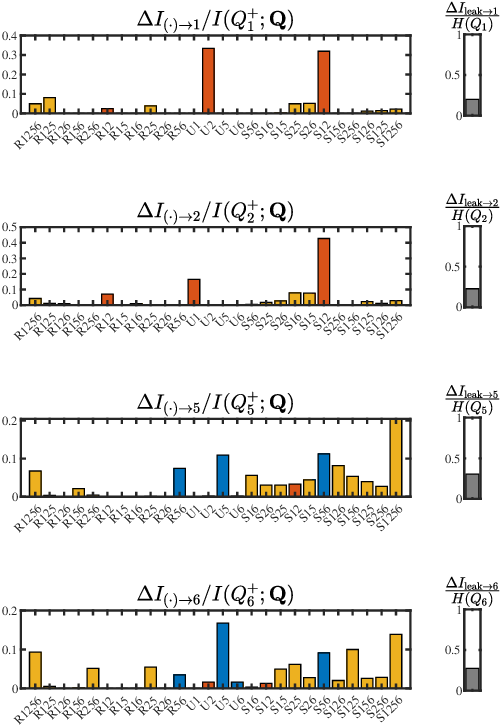}
  \end{center}
\caption{ Coupled R\"ossler-Lorenz system ($c=2$). The left panels
  show excerpts of the trajectories pertaining to R\"ossler systems
  $[Q_1,Q_2,Q_3]$ (top) and Lorenz system $[Q_4,Q_5,Q_6]$
  (bottom). The right panels show the redundant (R), unique (U), and
  synergistic (S) causalities among $[Q_1, Q_3, Q_4, Q_6]$. The
  information leak for each variable is also shown in the right-hand
  side bar.  The causalities are ordered from left to right according
  to $N_{\bi \rightarrow j}^{\alpha}$.  \label{fig:Rossler_c2}}
\end{figure}

The results for the coupled system ($c=2$) are shown in figure
\ref{fig:Rossler_c2}.  The left panel shows how the trajectory of the
Lorenz system is severely impacted by the coupling.  The new
causalities are shown in the right panel. As before, red and blue
colors are used to represent causalities exclusive to the R\"ossler
system (i.e., only involving $Q_1$ and $Q_2$) and Lorenz system (i.e.,
only involving $Q_5$ and $Q_6$), respectively, and yellow color is
used for causalities involving variables from both systems. The
causalities in the R\"ossler remain comparable to the uncoupled case
besides some small redundancies and synergies due to the effect of
unobserved variables. On the contrary, the causalities in the Lorenz
system undergo deeper changes. This is evidenced by the emergence of
multiple synergistic causalities involving $Q_1$ and $Q_2$. This
effect is consistent with the coupling of both systems. The emergence
of new redundant and synergistic causalities can be understood as a
more complex manifestation of the effect seen in the toy problem from
figure~\ref{fig:simple}: the combination of variables yields the
creation of redundancies synergies, where the latter dominate.
%

\section{Sensitivity of IT-causality to sample size and partition refinement}
\label{sec:convergence}

We investigate the sensitivity of IT-causality to the number of
samples ($N_{\text{samples}}$) used to estimate the probability
distributions and the number of bins employed to partition the range
of values of the variables ($N_{\text{bins}}$). The Lorenz system is
used as a testbed:
\begin{subequations}
\begin{align}  
  \label{eq:L}
 \frac{\dd Q_1}{\dd t} &= 10[ Q_2 - Q_1 ], \\
 \frac{\dd Q_2}{\dd t} &= Q_1 [28 - Q_3 ] - Q_2 \\
 \frac{\dd Q_3}{\dd t} &= Q_1 Q_2 - \frac{8}{3}Q_3.
\end{align}
\end{subequations}

The system was integrated over time to collect
$N_{\text{samples}}=5\times 10^3, 5\times 10^4, 5\times 10^5,$ and
$5\times10^8$ events after transients. Probability distributions were
calculated using uniform bins with $N_{\text{bins}}=10, 50, 100,$ and
200 per variable. Our primary focus is on causalities to $Q_1$, but
the conclusions drawn also apply to $Q_2$ and $Q_3$.

\begin{figure}
  \begin{center}
    \subfloat[]{\includegraphics[width=0.45\textwidth]{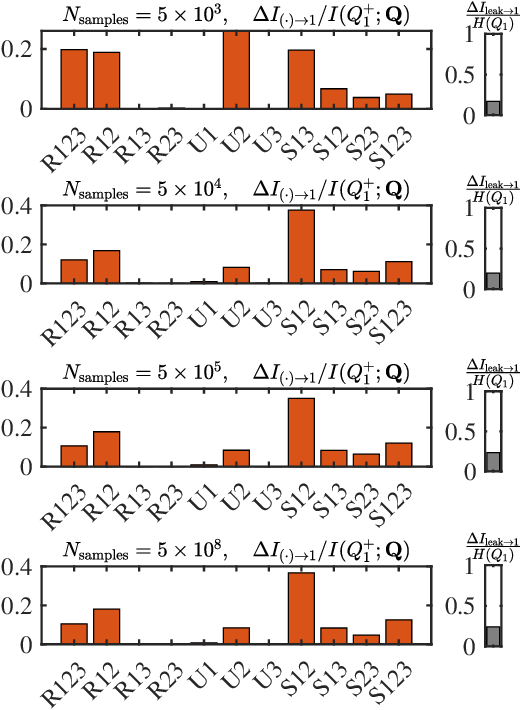}}
    \hspace{0.5cm}
    \subfloat[]{\includegraphics[width=0.45\textwidth]{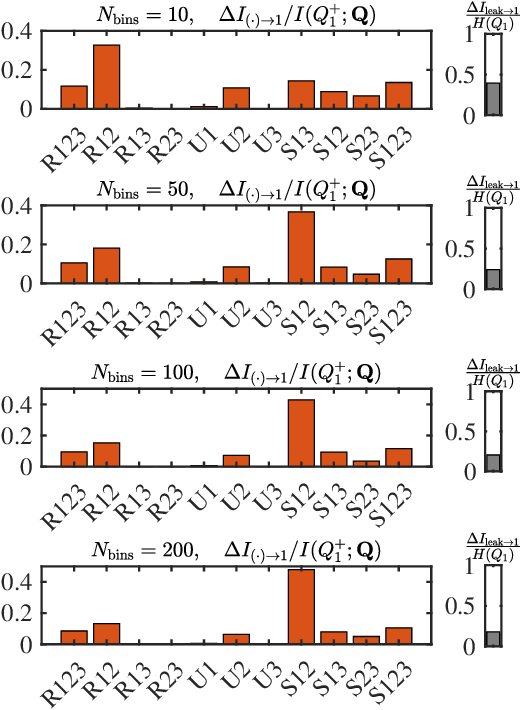}}
  \end{center}
\caption{ Sensitivity of IT-causality in the Lorenz system for (a)
  number of samples $N_\text{samples}$ used to estimate the
  probability distributions for $N_\text{bins}=50$ held constant and
  (b) the number of bins $N_\text{bins}$ used to partition values of
  the variables for $N_\text{samples}=10^5$. The causalities are
  ordered from left to right according to $N_{\bi \rightarrow
    1}^{\alpha}$.  \label{fig:covergence}}
\end{figure}

The sensitivity to $N_{\text{samples}}$ is displayed in
figure~\ref{fig:covergence}(a), where $N_{\text{samples}}$ varies
while maintaining $N_{\text{bins}}=50$ constant. For
$N_{\text{samples}}>5\times 10^3$, the changes in IT-causality remain
within a few percentage points of difference. The sensitivity to the
size of the partition is assessed in figure~\ref{fig:covergence}(b),
where $N_{\text{bins}}$ varies while $N_{\text{samples}}$ is held
constant at $N_{\text{samples}} = 5\times 10^5$. The IT-causalities
exhibit quantitative resemblance for all partitions, with the
exception of $N_{\text{bins}}=10$, which may be too coarse to capture
the continuous dynamics of the variables.
%

\bibliographystyle{jfm}
\bibliography{references}

\end{document}